\documentclass[preprint,aps]{revtex4}
\usepackage[dvipdfm]{graphicx}
\def\Vec#1{\mbox{\boldmath $#1$}}

\begin{document}

\title{
Single $\alpha$-particle orbits and Bose-Einstein condensation in $^{12}$C
}

\author{Taiichi Yamada}
\address{Laboratory of Physics, Kanto Gakuin University,
Yokohama 236-8501, Japan
}

\author{Peter Schuck}
\address{
\it Institute de Physique Nucl\'eaire, F-91406 Orsay Cedex, France
}

\date{\today}

\begin{abstract}

\vspace*{5mm}
\noindent
The $\alpha$-bosonic properties such as single-$\alpha$ orbits and occupation numbers
 in $J^\pi$=$0^+$, $2^+$, $1^-$ and $3^-$ states of $^{12}$C around the $3\alpha$ threshold 
 are investigated with the semi-microscopic $3\alpha$ cluster model.
As in other studies, 
 we found that the $0^+_2$ ($2^+_2$) state has dilute-$3\alpha$-condensate-like 
 structure in which the $\alpha$ particle is occupied in the single $S$ ($D$) orbit
 with about $70~\%$ ($80~\%$) probability.
The radial behaviors of the single-$\alpha$ orbits as well as the occupation numbers
 are discussed in detail in comparison with those for the $0^+_1$ and $2^+_1$ states
 together with the $1^-_1$ and $3^-_1$ states.
\\  
{PACS numbers: 21.10.Dr, 21.10.Gv, 21.60.Gx, 03.75.Hh}\\
\end{abstract}

\maketitle

\section{Introduction}

Four-nucleon correlations of the $\alpha$-cluster-type play an important role
 in nuclei.
The microscopic $\alpha$-cluster model~\cite{Wildermuth77,Brink66,Bertsch71,Fujiwara80}
 has succeeded in describing the structure of many states in light nuclei, in particular, 
 around the threshold energy of breakup into constituent clusters.
As for $^{12}$C, detailed analyses were 
 made by several authors with the microscopic $3\alpha$ cluster model about 25 years ago.
The $3\alpha$ GCM (generator coordinate method)~\cite{Uegaki77} and $3\alpha$ RGM 
 (resonating group method)~\cite{Fukushima77} calculations showed that 
 the second $0^+$ state of $^{12}$C ($E_x$=7.65 MeV), located at $E_{3\alpha}$=0.38 MeV 
 above the $3\alpha$ threshold, has a loosely coupled $3\alpha$ structure, 
 although the ground state is a shell-model-like compact state.
On the other hand, special attention has been paid to an $\alpha$-type condensation 
 in symmetric nuclear matter, analogue to the Bose-Einstein condensation 
 for finite number of dilute bosonic atoms such as $^{87}$Rb or $^{23}$Na 
 at very low temperature where all atoms occupy the lowest $S$ orbit~\cite{Dalfovo99}.
Several authors~\cite{Ropke98,Beyer00} showed the possibility of such $\alpha$-particle
 condensation in low-density nuclear matter, although the ordinary pairing 
 correlation can prevail at higher density.
This result suggests that dilute $\alpha$ condensate states in finite 
 nuclear system may exist in excited states as a weakly interacting
 gas of $\alpha$ particles.
  
Recently, a new $\alpha$-cluster wave function was proposed which is of
 the $N\alpha$ particle condensate type~\cite{Tohsaki01}
\begin{eqnarray}
&&|\Phi_{N\alpha}\rangle=(C^+_\alpha)^N |{\rm vac}\rangle,\label{C_dager}\\
&&\langle\Vec{r}_1\cdots\Vec{r}_N|\Phi_{N\alpha}\rangle\propto{\mathcal A}\left\{ e^{-\nu\left(\Vec{r}_1^2+\cdots+\Vec{r}_N^2\right)}
\phi(\alpha_1)\cdots\phi(\alpha_N)\right\},\label{schuck_type_wf}
\end{eqnarray}
 where $C^+_\alpha$ is the $\alpha$-particle creation operator, $\phi(\alpha)$ denotes
 the internal wave function of the $\alpha$ cluster, $\Vec{r}_i$ is the center-of-mass
 coordinate of the {\it i}-th $\alpha$ cluster, and ${\mathcal A}$ presents 
 the antisymmetrizer among the nucleons belonging to different $\alpha$ clusters.
The important characteristic of the wave function is that the center-of-mass
 motion of each $\alpha$ cluster is of $S$-wave type.
Applications of the condensate-type wave function to $^{12}$C and $^{16}$O~\cite{Tohsaki01}
 indicated that the second $0^+$ state of $^{12}$C ($E_x$=7.65 MeV) and
 fifth $0^+$ state of $^{16}$O ($E_x$=14.0 MeV), around the $3\alpha$ and
 $4\alpha$ threshold, respectively, are conjectured to be dilute $N\alpha$
 condensate states, which are quite similar to the Bose-Einstein condensation
 of bosonic atoms at very low temperature.
The calculated nuclear radii for both of those states are about 4 fm,
 significantly larger than that for the ground state (about 2.5 fm).
As for $^{12}$C, a detailed analysis with a deformed alpha condensate
 wave function, slightly different from the spherical one
 in Eq.~(\ref{schuck_type_wf}), was performed to investigate 
 the structure of the $0^+_1$ and $0^+_2$ states.~\cite{Funaki02}~
It was found that each of the $0^+_2$ wave functions obtained
 by the $3\alpha$ GCM and RGM calculations has a large squared overlap value
 of more than $90 \%$ with the single $3\alpha$ condensate wave function.
 
The above-mentioned results for $^{12}$C and $^{16}$O lead us to the further
 intriguing issue that dilute $\alpha$-cluster states with $J^\pi=0^+$ near the $N\alpha$
 threshold may exist in other heavier $4N$ self-conjugate nuclei.
The Gross-Pitaevskii-equation approach~\cite{Yamada04} is useful to explore 
 such dilute multi $\alpha$ systems, because this equation~\cite{Pitaevskii61}, based on
 mean field theory, has succeeded in describing
 the structure of the Bose-Einstein condensation for dilute neutral 
 atomic systems, for example, $^{87}$Rb or $^{23}$Na, at very low temperature,
 trapped by an external magnetic field.~\cite{Dalfovo99}~
The present authors~\cite{Yamada04} applied the Gross-Pitaevskii equation to self-conjugate
 $4N$ nuclei.
They found that 1)~there exists a critical number of $\alpha$ bosons
 that the dilute $N\alpha$ system can sustain as a self-bound nucleus, and
 2)~the Coulomb-potential barrier plays an important role to confine such 
 dilute $N\alpha$-particle condensate states.  

It is interesting to explore also the possibility of the $\alpha$ condensate states with non-zero
 angular momentum in $^{12}$C.
The old theoretical calculations based on the microscopic $3\alpha$ cluster 
 model~\cite{Fujiwara80,Uegaki77,Fukushima77} suggested 
 the existence of a $2^+_2$ state of $^{12}$C at around $E_{3\alpha}\sim3$ MeV above the $3\alpha$
 threshold, the structure of which is similar to the $0^+_2$ state except for the angular momentum.
Quite recently the $2^+_2$ state was observed at $E_{3\alpha}=2.6\pm0.3$ MeV with the alpha decay
 width $\Gamma_\alpha=1.0\pm0.3$ MeV.~\cite{Itoh04}~
The $\alpha$ condensate-type wave function with axially deformation~\cite{Funaki04} was applied to
 study the structure of the $2^+_2$ state with help of the method of ACCC
 (analytic continuation in the coupling constant)~\cite{Kukulin77}.
They found that the $2^+_2$ state has a large overlap with the single
 condensate wave function of $3\alpha$ gas-like structure, the squared value of
 which amounts to about $88 \%$.
This result implies that the $2^+_2$ state has a similar structure as the $0^+_2$ state,
 namely, dilute $3\alpha$ condensation.

Here, it is an intriguing problem to discuss whether the $0^+_2$ and $2^+_2$ states
 of $^{12}$C are {\it ideal} dilute $\alpha$ condensates or not.  
The condensate-type $\alpha$-cluster wave function in Eq.~(\ref{schuck_type_wf}) 
 has succeeded in describing the $0^+_2$ state of $^{12}$C.
This result, however, does not necessarily mean that the $0^+_2$ state of $^{12}$C is 
 an ideal $\alpha$-condensate state.
If the $0^+_2$ state of $^{12}$C is an ideal dilute $\alpha$-condensate, 
 the single $\alpha$-particle orbit in the state should be of the zero-node 
 long-ranged $S$-wave type with an occupation probability of $100~\%$, as suggested
 from the Gross-Pitaevskii-equation approach~\cite{Yamada04}.
The antisymmetrizer ${\mathcal A}$ in Eq.~(\ref{schuck_type_wf}) generally perturbs
 the single $\alpha$ motion in the nucleus, and one should remember that
 the condensate-type wave function 
 can also describe the shell-model-like compact structure of the ground state of $^{12}$C.
The effect of the antisymmetrizer should have a close relation to the rms radius of the nucleus 
 or the distance between $2\alpha$ clusters in a nucleus. 
The ideal $3\alpha$ condensate state is expected  to be realized if the distance 
 between two arbitrary $\alpha$ clusters is large enough so that the effect of the antisymmetrizer 
 can be neglected.
The calculated nuclear radius for the $0^+_2$ state of $^{12}$C, about 4 fm~\cite{Tohsaki01}, 
 suggests that the action of the antisymmetrizer is weakened significantly in that state.
In order to give more decisive theoretical evidence that the $0^+_2$ state
 of $^{12}$C as well as the $2^+_2$ state has dilute $3\alpha$ condensation structure, 
 it is needed to study quantitatively the bosonic properties such as single $\alpha$-particle 
 orbits and corresponding occupation probabilities, starting from the microscopic wave function.

The first attempt to derive the $\alpha$-boson properties for $0^+$ states in $^{12}$C 
 from a microscopic model was performed in Ref.~\cite{Matsuura04}, 
 where the $3\alpha$ RGM equation was solved in terms of the correlated Gaussian basis
 with the stochastic variational method. 
Although they formulated a derivation of the $3\alpha$ boson wave function starting from the microscopic 
 $3\alpha$ wave function, the $\alpha$ bosonic properties of $^{12}$C were studied not
 with the $3\alpha$ bosonic wave function but with the normalized spectroscopic amplitude, 
 because the derivation of the $3\alpha$ boson wave function is numerically difficult due to 
 the non-local properties of the norm kernel.
Although the normalized spectroscopic amplitude seems to be a good approximation 
 for the boson wave function in the region where the effect of the antisymmetrizer is negligible, 
 the approximation becomes worse when the spatial overlap of the $3\alpha$ clusters becomes larger. 
It is requested to demonstrate quantitatively how good the approximation is 
 for the $0^+_2$ state within their framework.

The purposes in the present paper are twofold.
First we study the bosonic properties such as single 
 $\alpha$-particle orbits and occupation probabilities for the $0^+$ and $2^+$ states in $^{12}$C
 with direct use of the wave function obtained by the $3\alpha$ OCM (orthogonality condition model)~\cite{Saito68}.
The OCM is a semi-microscopic model and a simple version of the RGM,
 taking into account properly the antisymmetrization among nucleons,
 which successfully describes the structure of $^{12}$C.~\cite{Fujiwara80,Horiuchi74,Kato89,Kurokawa04}~
The second purpose is to explore the possibility of the dilute $3\alpha$ condensation with
 {\it negative} parity within the present framework.
The $3^-_1$ ($1^-_1$) state of $^{12}$C at $E_{3\alpha}=2.37$ (3.57) MeV 
 above the $3\alpha$ threshold appears at the same energy region as 
 that for the $0^+_2$ ($E_x$=0.38 MeV) and $2^+_2$ (2.6 MeV) states.
According to the old theoretical study based on the $3\alpha$ GCM and RGM 
 calculations~\cite{Uegaki77,Fukushima77},
 the nuclear radius of the $3^-$ state is intermediate between a compact shell-model-like
 state ($0^+_1$) and a loosely coupled $3\alpha$ cluster state ($0^+_2$), while
 the $1^-$ state has a radius only a little smaller than that of the $0^+_2$ state.  
Thus, it is quite interesting to study the bosonic properties for the negative
 parity states, as well.

In order to clarify the relationship between the $N\alpha$ boson
 wave function and the one of $N\alpha$ OCM, we first outline a way of mapping 
 of the microscopic $N\alpha$ wave function onto 
 the $N\alpha$ boson wave function, and derive the equation of motion for $N\alpha$ bosons 
 from the $N\alpha$ RGM equation in Sec.~II.
The $N\alpha$ OCM equation is illustrated as an approximation of the $N\alpha$ boson
 equation.
The $N\alpha$ OCM wave function, thus, has bosonic properties.
The $3\alpha$ OCM equation is solved properly with modern numerical techniques. 
The calculated single $\alpha$-particle orbits and occupation probabilities in $^{12}$C 
 are discussed in Sec.~III.
Finally, we give the summary in Sec.~IV.

\newpage
\section{Formulation}

A way of mapping of the microscopic $N\alpha$ wave function onto 
 the $N\alpha$ boson wave function is illustrated in order to derive the equation of motion for 
 the $N\alpha$ bosons. 
The $N\alpha$ OCM equation is given as an approximation of the equation of motion for
 $N\alpha$ bosons.
We formulate the evaluation of the single-$\alpha$ orbits and occupation numbers
 from the $N\alpha$ OCM wave function together with other physical quantities.
Finally, we give an outline of how to solve the $3\alpha$ OCM equation for $^{12}$C
 with a phenomenological $\alpha\alpha$ potential.

\subsection{Mapping of the fermionic $N\alpha$ wave function onto a $N\alpha$ boson wave function}

In the microscopic $N\alpha$ cluster model, the total wave function, $\Psi_J^{(F)}$, 
with the total angular momentum $J$ is given as
\begin{eqnarray}
\Psi_J^{(F)}&=&\mathcal{A}\left\{\prod_{i=1}^{N}\phi^{(int)}_{\alpha_i}\chi_J\left(\Vec{r}\right)\right\},\label{total_wf_1}\\
       &=&\int d\Vec{a} \Psi_J^{(F)}(\Vec{a}) \chi (\Vec{a}), \label{total_wf_2}
\end{eqnarray}  
where $\phi^{(int)}_\alpha$ denotes the intrinsic wave function of the $\alpha$ particle 
 with the simple $(0s)^4$ shell-model configuration, and $\chi_J$ represents 
 the relative wave function with a set of the relative (Jacobi) coordinates, 
 $\Vec{r}=\{\Vec{r}_1, \Vec{r}_2,\cdots, \Vec{r}_{N-1}\}$, with respect to the c.m. of $\alpha$ clusters. 
The antisymmetrization among $4N$ nucleons is properly taken into account in terms of 
 the operator $\mathcal{A}$.
The function $\Psi_J^{(F)}(\Vec{a})$ in Eq.~(\ref{total_wf_2}) is defined as
\begin{eqnarray}
\Psi_J^{(F)}(\Vec{a})=\mathcal{A}\left\{\prod_{i=1}^{N}\phi^{(int)}_{\alpha_i}\prod_{j=1}^{N-1}\delta(\Vec{r}_j-\Vec{a}_j)\right\},      
\end{eqnarray}
which describes the $\alpha$-cluster state located at the relative positions
 specified by the Jacobi parameter coordinate $\Vec{a}=\{\Vec{a}_1, \Vec{a}_2, \cdots, \Vec{a}_{N-1}\}$.

The total Hamiltonian for $4N$ fermions is given as
\begin{eqnarray}
H=\sum_{i=1}^{4N} t_i-T_{cm}+\sum_{i<j=1}^{4N}\left(\upsilon_{ij}+\upsilon_{ij}^{Coul}\right),
\end{eqnarray} 
where $t_i$ and $T_{cm}$ denote, respectively, the kinetic energy operator of the {\it i}-th nucleon
 and of the center-of-mass of the total system.
The nucleon-nucleon interaction (Coulomb interaction) between the {\it i}-th and {\it j}-th nucleons
 is expressed as $\upsilon_{ij}$ ($\upsilon_{ij}^{Coul}$).

The Schr\"odinger equation for the fermionic $N\alpha$ system is 
\begin{eqnarray}
H\Psi_J^{(F)}=E\Psi_J^{(F)}. \label{schrodinger_eq}
\end{eqnarray}
Substituting the total wave function of Eq.~(\ref{total_wf_2}) into Eq.~(\ref{schrodinger_eq}), 
 we obtain the equation of motion for the relative wave function $\chi_J$, 
\begin{eqnarray}
\int d\Vec{a}' \left\{ H(\Vec{a}, \Vec{a}') - E N(\Vec{a}, \Vec{a}') \right\} \chi_J (\Vec{a}')=0,
  \hspace*{5mm}{\rm or}\hspace*{5mm}
  \left(\mathcal{H} - \mathcal{N}\right)\chi_J=0, \label{rgm_eq}
\end{eqnarray}
which is called the RGM (resonating group method) equation~\cite{Horiuchi86}.
The Hamiltonian and norm kernels, $H(\Vec{a},\Vec{a}')$ and $N(\Vec{a},\Vec{a}')$,
 are defined as
\begin{eqnarray}
\left\{ \begin{array}{c} H(\Vec{a}, \Vec{a}') \\ N(\Vec{a}, \Vec{a}') \end{array} \right\}
  = {\langle \Psi_J^{(F)}(\Vec{a}) \mid \left\{ \begin{array}{c} H \\ 1 \end{array} \right\}
      \mid \Psi_J^{(F)}(\Vec{a}')\rangle}.
\end{eqnarray}  

Recalling the normalization condition ${\langle \Psi_J^{(F)}\mid \Psi_J^{(F)}\rangle=1}$ 
 for the total wave function in Eq.~(\ref{total_wf_1}),
 the normalization of $\chi_J$ in Eq.~(\ref{rgm_eq}) is given by
\begin{eqnarray}
\int d\Vec{a} d\Vec{a}' \chi_J^*(\Vec{a}') N(\Vec{a},\Vec{a}') \chi_J(\Vec{a}')=1.
\end{eqnarray}
This suggests that an $N\alpha$ boson wave function $\Phi_J^{(B)}$ corresponding to the fermionic 
 wave function $\Psi_J^{(F)}$ in Eq.~(\ref{total_wf_1}) should be taken to be
\begin{eqnarray}
\Phi_J^{(B)}(\Vec{a})=\mathcal{N}^{1/2}\chi_J=\int d\Vec{a}'{N}^{1/2}(\Vec{a},\Vec{a}')\chi_J(\Vec{a}'),\label{boson_wf}
\hspace*{5mm} \int d\Vec{a} {\left| \Phi_J^{(B)}(\Vec{a}) \right|^2}=1,
\end{eqnarray}
where ${N}^{1/2}$ is defined as
\begin{eqnarray}
\int d\Vec{a}''{N}^{1/2}(\Vec{a},\Vec{a}'') {N}^{1/2}(\Vec{a}'',\Vec{a}')={N}(\Vec{a},\Vec{a}').
\end{eqnarray}
It is noted that the boson wave function $\Phi_J^{(B)}(\Vec{a})$ has only the Jacobi
 coordinates $\Vec{a}=\{\Vec{a}_1,\Vec{a}_2,\cdots,\Vec{a}_{N-1}\}$ of the $N\alpha$ system and
 the internal coordinates in the $\alpha$ cluster are integrated out completely.
In addition, $\Phi_J^{(B)}(\Vec{a})$ is totally symmetric for any two-$\alpha$-cluster exchange.
Thus, it has bosonic property.
From the RGM equation~(\ref{rgm_eq}), $\Phi_J^{(B)}$ should satisfy the following equation
\begin{eqnarray}
\left({\mathcal{N}^{-1/2}}\mathcal{H}{\mathcal{N}^{-1/2}}-E\right)\Phi_J^{(B)}=0, \label{boson_eq}
\end{eqnarray} 
Thus, we can interpret $\mathcal{N}^{-1/2}\mathcal{H}\mathcal{N}^{-1/2}$ as the $N\alpha$ boson Hamiltonian,
 and Equation~(\ref{boson_eq}) is the equation of motion for the $N\alpha$ boson wave function.

If the eigenvalue problem for the norm kernel
\begin{eqnarray}
\mathcal{N}u_\lambda=\int d\Vec{a}' N(\Vec{a},\Vec{a}') u_\lambda(\Vec{a}')=\lambda u_\lambda(\Vec{a}) 
       \label{eigenvalue_eq}
\end{eqnarray}
is solved, the boson wave function $\Phi_J^{(B)}(\Vec{a})$ in Eq.~(\ref{boson_wf}), then, is obtained as
\begin{eqnarray}
\Phi_J^{(B)}(\Vec{a})=\sum_\lambda \sqrt{\lambda} u_\lambda(\Vec{a}) {\langle u_\lambda\mid\chi_J\rangle}.
        \label{boson_wf_2}
\end{eqnarray}
The eigenvalue of the norm kernel, $\lambda$, is non-negative, and the eigenfunction $u_\lambda$ 
 with $\lambda=0$ is called the Pauli-forbidden state, which satisfies the condition 
 $\mathcal{A}\{\prod_{i=1}^N\phi^{(int)}_{\alpha_i} u_\lambda\}=0$.
The boson wave function $\Phi_J^{(B)}$, thus, has no component of the Pauli-forbidden state, 
\begin{eqnarray}
 {\langle u_\lambda \mid \Phi_J^{(B)}\rangle=0} 
\hspace*{5mm}{\rm for}\hspace*{3mm}u_\lambda\hspace*{3mm}{\rm with}\hspace*{3mm} \lambda=0.
\label{orthogonality_condition_RGM}
\end{eqnarray}

\subsection{$N\alpha$ orthogonality condition model (OCM) with bosonic properties}

In the previous section, we mapped the fermionic wave function onto the $N\alpha$ boson wave function
 $\Phi_J^{(B)}$ within the framework of the resonating group method (RGM).
The boson wave function has the following properties: 1) $\Phi_J^{(B)}$ is totally symmetric for 
 any 2$\alpha$-particle exchange, 2) $\Phi_J^{(B)}$ satisfies the equation motion in Eq.~(\ref{boson_eq}),
 and 3) $\Phi_J^{(B)}$ is orthogonal to the Pauli forbidden state (see Eq.~(\ref{orthogonality_condition_RGM})).
In order to obtain the boson wave function, we need to solve the RGM equation~(\ref{rgm_eq})
 and the eigenvalue equation of the norm kernel~(\ref{eigenvalue_eq}) or to solve directly
 the equation of motion~(\ref{boson_eq}).
Solving the eigenvalue equation of the norm kernel, however, is difficult in general even for the 3$\alpha$ case.
In addition, computational problems are encountered for solving the $N\alpha$ RGM 
 equation~(\ref{rgm_eq}) for $N \geq 4$.
Thus, it is requested to use more feasible frameworks for the study of the bosonic properties and
 the amount of $\alpha$ condensation for the $N\alpha$ system.  
In the present study, we take the orthogonality condition model (OCM)~\cite{Saito68} 
 as one of the more feasible frameworks.
The OCM scheme, which  is an approximation to the RGM, is known to describe nicely
 the structure of low-lying states in light nuclei~\cite{Fujiwara80,Saito68,Horiuchi74,Kato89,Kurokawa04}.
The essential properties of the $N\alpha$ boson wave function $\Phi_J^{(B)}$, as mentioned above,
 can be taken into account in OCM in a simple manner.
We will demonstrate this briefly.
 
In OCM, the $\alpha$ cluster is treated as a point-like particle.
We approximate the $N\alpha$ boson Hamiltonian (non-local) in Eq.~(\ref{boson_eq}) 
 by an effective (local) one $H^{{\rm (OCM)}}$,
\begin{eqnarray}
  && {\mathcal{N}^{-1/2}}\mathcal{H}{\mathcal{N}^{-1/2}} \sim H^{{\rm (OCM)}} \\
  && H^{{\rm (OCM)}}\equiv \sum_{i=1}^N T_i - T_{cm}+\sum_{i<j=1}^{N} V_{2\alpha}^{eff}(i,j)+\sum_{i<j<k=1}^{N}V_{3\alpha}^{eff}(i,j,k),
      \label{hamiltonian_ocm}
\end{eqnarray}
where $T_i$ denotes the kinetic energy of the {\it i}-th $\alpha$ cluster, and the center-of-mass kinetic
 energy $T_{cm}$ is subtracted from the Hamiltonian.
The effective 2$\alpha$ and 3$\alpha$ potentials are presented as $V_{2\alpha}^{eff}$ and $V_{3\alpha}^{eff}$,
 respectively.
Referring to the RGM framework in Eqs.~(\ref{boson_eq})$\sim$(\ref{orthogonality_condition_RGM}),
  the equation of the relative motions for the $N\alpha$ particles with $H^{{\rm (OCM)}}$, 
  called the OCM equation, is expressed as
\begin{eqnarray}
 &&\left\{ H^{{\rm (OCM)}}-E \right\}\Phi_J=0, \label{ocm_eq} \\
 &&{\langle u_F \mid \Phi_J \rangle}=0, \label{orthogonality_condition}
\end{eqnarray}
where $u_F$ denotes the Pauli-forbidden state, satisfying the following condition
\begin{eqnarray} 
  \mathcal{A}\{\prod_{i=1}^N\phi^{(int)}_{\alpha_i} u_F \}=0.
\end{eqnarray}
The orthogonality condition in Eq.~(\ref{orthogonality_condition}) corresponds 
 to Eq.~(\ref{orthogonality_condition_RGM}) in Sec.~IIA.
The bosonic property of the wave function $\Phi_J$ can be taken into account
 by symmetrizing the wave function with respect to any 2$\alpha$-particle exchange,
\begin{eqnarray}
\Phi_J=\mathcal{S}\Phi_J(1,2,\cdots,N),
\end{eqnarray}
where $\mathcal{S}$ denotes the symmetrization operator, 
 $\mathcal{S}=(1/\sqrt{N!})\sum_\mathcal{P} \mathcal{P}$, where the sum runs over 
 all permutations $\mathcal{P}$ for the $N\alpha$ particles.
It is noted that the complete overlapped state of the $3\alpha$ particles is forbidden 
 within the present framework because of the Pauli-blocking effect in Eq.~(\ref{orthogonality_condition}), 
 although we take into account the bosonic statistics for the constituent $\alpha$ particles.
In the next section, we will demonstrate i)~how to solve the OCM equation 
 and ii)~what kind of effective $\alpha$-$\alpha$ potential we should choose in $H^{{\rm (OCM)}}$ 
 for the $3\alpha$ case of $^{12}$C.

Here, it is useful to define various quantities characterizing the structure of the $N\alpha$
 system with use of the $N\alpha$ boson wave function $\Phi_J$ obtained by solving the OCM equation
 in Eqs.~(\ref{ocm_eq}) and (\ref{orthogonality_condition}).
The single $\alpha$-particle density is defined as
\begin{eqnarray}
\rho(\Vec{r})=\langle\Phi_J\mid \sum_{i=1}^N \delta(\Vec{r}-\Vec{r}_i^{(cm)}) \mid\Phi_J\rangle,
\label{single_density}
\end{eqnarray}
where $\Vec{r}_i^{(cm)}$ is the coordinate of the {\it i}-th $\alpha$ particle measured
 from the center-of-mass coordinate of the total system.
The nuclear root-mean-square radius is given as
\begin{eqnarray}
\sqrt{{\langle r^2_N\rangle}}=\sqrt{{\langle r^2_\alpha\rangle}+1.71^2},\\
\sqrt{{\langle r^2_\alpha\rangle}}=\int d\Vec{r} r^2\rho(\Vec{r}),
\end{eqnarray}
where we take into account the finite size effect of the $\alpha$ particle.
The correlation functions with respect to the $\alpha$-$\alpha$ relative coordinate $\Vec{r}_{\alpha\alpha}$
 as well as the relative coordinate between one of the $\alpha$ particles and 
 the remaining $(N-1)\alpha$ system $\Vec{r}_{\alpha-(N-1)\alpha}$ are given as 
\begin{eqnarray}
&&f_{\alpha\alpha}(r)=\langle\Phi_J \mid \delta(\Vec{r}-\Vec{r}_{\alpha\alpha}) \mid\Phi_J\rangle,\label{cf_2alpha}\\
&&f_{\alpha -{(N-1)\alpha}}(r)=\langle\Phi_J \mid \delta(\Vec{r}-\Vec{r}_{\alpha - {(N-1)\alpha}}) \mid\Phi_J\rangle,
\label{cf_alpha_2alpha}
\end{eqnarray}
where the way of choosing the coordinates, $\Vec{r}_{\alpha\alpha}$ and $\Vec{r}_{\alpha-(N-1)\alpha}$, 
 is arbitrary in the set of Jacobi coordinates of the $N\alpha$ particles because of the totally symmetrization for $\Phi_J$. 
The root-mean-square (rms) distances with respect to $\Vec{r}_{\alpha\alpha}$ and $\Vec{r}_{\alpha-(N-1)\alpha}$
are, respectively, given by
\begin{eqnarray}
&&\sqrt{\langle r_{\alpha\alpha}^2\rangle}=\left[\langle \Phi_J \mid \Vec{r}_{\alpha\alpha}^2 \mid \Phi_J \rangle \right]^{1/2},\\
&&\sqrt{\langle r_{\alpha-{(N-1)\alpha}}^2\rangle}=\left[\langle \Phi_J \mid \Vec{r}_{\alpha-{(N-1)\alpha}}^2 \mid \Phi_J \rangle \right]^{1/2}.
\end{eqnarray} 
The reduced width amplitude for the $\alpha$-$(N-1)\alpha$ part is defined as
\begin{eqnarray}
\mathcal{Y}_{\ell L J}(r)=r\times
\left\langle\left[Y_L(\Vec{r})
\phi_\ell\left((N-1)\alpha\right)\right]_J\mid \Phi_J\right\rangle, \label{rwa}
\end{eqnarray}
where \Vec{r} denotes the relative coordinate between the $\alpha$ particle and the $(N-1)\alpha$ nucleus, and
 $\phi_\ell\left((N-1)\alpha\right)$ represents the wave function of the $(N-1)\alpha$ nucleus
 with total angular momentum $\ell$ which is obtained by solving the OCM equation
 for the $(N-1)\alpha$ system.
The integration in Eq.~(\ref{rwa}) is done over all of the relative (Jacobi) coordinates for
 the $N\alpha$ system except for the radial part of $\Vec{r}$.

In order to discuss the bosonic properties such as the degree of $\alpha$ condensation in a nucleus,
 it is needed to extract information on the single $\alpha$-particle orbits
 and its occupation probabilities in the nucleus from the total wave function $\Phi_J$.
The one-particle density matrix for the $N\alpha$ system is very useful for this~\cite{Matsuura04}.
Defining the one-particle density operator as
\begin{eqnarray}
\mathcal{D}(\Vec{r},\Vec{r}')={\sum_{i=1}^N \mid \delta(\Vec{r_i^{(cm)}}-\Vec{r}')\rangle\langle\delta(\Vec{r_i^{(cm)}}-\Vec{r}) \mid},
\end{eqnarray}
then, the single $\alpha$-particle density matrix is given as
\begin{eqnarray}
\rho(\Vec{r},\Vec{r}') &=&
\langle\Phi_J\mid \mathcal{D}(\Vec{r},\Vec{r}') \mid\Phi_J\rangle,\\
&=& N\times \langle\Phi_J\mid \delta(\Vec{r_1^{(cm)}}-\Vec{r}')\rangle\langle\delta(\Vec{r_1^{(cm)}}-\Vec{r}) \mid\Phi_J\rangle,
\label{one_body_density}
\end{eqnarray}
where $\Vec{r}_i^{(cm)}$ is the same as that in Eq.~(\ref{single_density}).
It is noted that the diagonal matrix element reduces to the single $\alpha$-particle density
 in Eq.~(\ref{single_density}):~$\rho(\Vec{r},\Vec{r}'=\Vec{r})=\rho(\Vec{r})$.
The single $\alpha$-particle orbit and its occupation number in the nucleus
 can be evaluated by solving the eigenvalue equation of the one-particle density matrix
\begin{eqnarray}
\int d\Vec{r}'\rho(\Vec{r},\Vec{r}')\varphi_\mu(\Vec{r}')=\mu\varphi_\mu(\Vec{r}),\label{eigenvalue_equation_one_body_density}
\end{eqnarray}
where the eigenvalue $\mu$ presents the occupation number.
The eigenfunction $\varphi_\mu$ denotes the single-$\alpha$ orbital wave function in a nucleus 
 with the argument of the intrinsic coordinate ($\Vec{r}^{(cm)}_\alpha$) of an arbitrary $\alpha$ particle in a nucleus
 measured from the center-of-mass coordinate.
The ratio $\mu/N$ represents the occupation probability of an $\alpha$ particle 
 in the orbit $\varphi_\mu$.
The spectrum of the occupation probabilities offers important information about the occupancy of
 the single $\alpha$-particle orbit in a nucleus.
If each of the $N\alpha$ particles in an $N\alpha$-boson state is occupied
 by only one orbit, the occupation probability for this orbit becomes $100 \%$. 

The $^8$Be ($2\alpha$) system is a good example to demonstrate the characteristic
 of the single-$\alpha$ orbital wave function.
From the definition of Eqs.~(\ref{one_body_density}) and (\ref{eigenvalue_equation_one_body_density}), 
 the $L_\alpha$-wave single-$\alpha$ orbit in the $^8$Be($J^\pi$) state 
 corresponds to the relative wave function (which is obtained by solving the $2\alpha$ OCM 
 equation with $J$ ($=L_\alpha$) in Eqs.~(\ref{ocm_eq}) and (\ref{orthogonality_condition})), 
 scaling to $1/2$ with respect to the relative coordinate between the $2\alpha$ clusters.
Then, the occupation probability becomes exactly (mathematically) $100 \%$ for any $L_\alpha$-value.

The radial behavior of the $L_\alpha$-wave single-$\alpha$ orbit, 
 $\varphi_\mu({r}^{(cm)}_\alpha)$, in Eq.~(\ref{eigenvalue_equation_one_body_density}) generally has a close
 relationship with that of the reduced width amplitude, $\mathcal{Y}_{\ell L J}(r_{\alpha-(N-1)\alpha})$, 
 in Eq.~(\ref{rwa}).
This is due to the fact that both represent the behavior of the single-$\alpha$-particle motion
 in a nucleus in which all degrees of freedom of the other $\alpha$ particles are integrated out, 
 and $\Vec{r}^{(cm)}_\alpha$ is given as $\Vec{r}^{(cm)}_\alpha=\frac{N-1}{N}\times\Vec{r}_{\alpha-(N-1)\alpha}$. 

The momentum distribution of the single $\alpha$ particle is also helpful for the study of 
 $\alpha$ condensation in a nucleus~\cite{Matsuura04}.
It is defined as a double Fourier transformation of the one-particle density matrix
\begin{eqnarray}
&&\rho(k)=\int d\Vec{r}' d\Vec{r} \frac{e^{i\Vec{k}\cdot\Vec{r}'}}{(2\pi)^{3/2}}
             \rho(\Vec{r},\Vec{r}')\frac{e^{-i\Vec{k}\cdot\Vec{r}}}{(2\pi)^{3/2}},\\
&&\int d\Vec{k}\rho(k)=1,
\end{eqnarray}
It is reminded that $\rho(k)$ would have a $\delta$-function like peak around $k=0$
 for an ideal dilute condensed state of infinite size.

\subsection{3$\alpha$ OCM for $^{12}$C}

In the previous section, we outlined the $N\alpha$ orthogonality condition model (OCM)
 and discussed how to extract the properties and the amount of $\alpha$ condensation
 in the $N\alpha$ system.
Here, we will apply the OCM framework to the $3\alpha$ system of $^{12}$C.

\begin{figure}
\begin{center}
\includegraphics*[scale=0.8,clip]{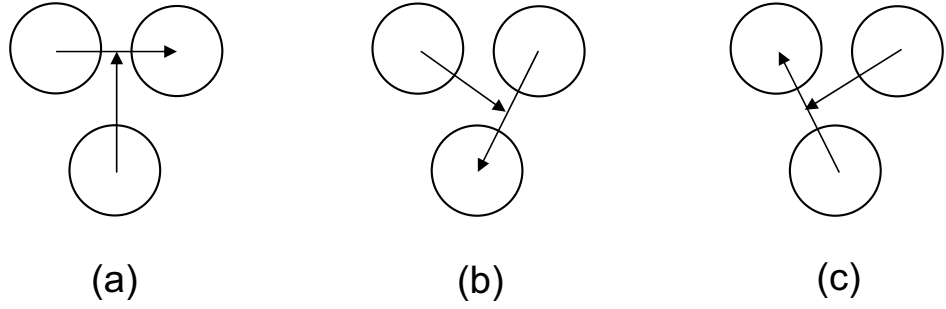}
\caption{
Jacobian coordinate systems for the $3\alpha$ model of $^{12}$C.
The three Jacobian coordinates, (a), (b) and (c),
 correspond respectively to the $3\alpha$ relative wave functions, 
 $\Phi_J^{3\alpha}(12,3)$, $\Phi_J^{3\alpha}(23,1)$ and $\Phi_J^{3\alpha}(31,2)$
 in Eq.~(\ref{3alpha_total_wf}).
}
\label{fig:1}
\end{center}
\end{figure}

The total wave function of $^{12}$C (the total angular momentum $J$) within the frame of 
 the $3\alpha$ OCM is presented as
\begin{eqnarray}
\Phi_J({^{12}{\rm C}})=\Phi_J^{(12,3)}+\Phi_J^{(23,1)}+\Phi_J^{(31,2)},\label{3alpha_total_wf}
\end{eqnarray}
where $\Phi_J^{(12,3)}$ denotes the relative wave function of the $3\alpha$ system
 with the Jacobi-coordinate system shown in Fig.~\ref{fig:1}(a), and others are self-explanatory.
In the present study, $\Phi_J({^{12}{\rm C}})$ is expanded in terms of the Gaussian basis~\cite{Kamimura88},
\begin{eqnarray}
&&\Phi_J({^{12}{\rm C}})=\sum_c\sum_{\nu,\mu}A_c(\nu,\mu)\Phi_c^{3\alpha}(\nu,\mu),\label{gaussian_basis}\\
&&\Phi_c^{3\alpha}(\nu,\mu)=\Phi_c^{(12,3)}(\nu,\mu)+\Phi_c^{(23,1)}(\nu,\mu)+\Phi_c^{(31,2)}(\nu,\mu),\label{3alpha_OCM_basis}\\
&&\Phi_c^{(ij,k)}(\nu,\mu)=\left[\varphi_{\ell}(\Vec{r}_{ij},\nu)\varphi_{L}(\Vec{r}_k,\mu)\right]_J,\\
&&\varphi_\ell(\Vec{r},\nu)=N_\ell(\nu)r^\ell \exp(-\nu r^2) Y_\ell (\hat{\Vec{r}}), 
\end{eqnarray}
where $N_\ell$ is the normalization factor, and $\Vec{r}_{ij}$ ($\Vec{r}_k$) denotes
 the relative coordinate between the {\it i}- and {\it j}-th $\alpha$ particle
 (the {\it k}-th $\alpha$ particle and the center-of-mass coordinate of the {\it i}-th
 and {\it j}-th $\alpha$ particle).   
The angular momentum channel is presented as $c=(\ell,L)_J$, where $\ell$ ($L$) denotes 
 the relative orbital angular momentum between 2$\alpha$ clusters 
 (the center-of-mass for the $2\alpha$ clusters and the third $\alpha$).
The Gaussian parameter $\nu$ is taken to be of geometrical progression,
\begin{equation}
\nu_n=1/b_n^2,\hspace{1cm}b_n=b_{\rm min}a^{n-1},\hspace{1cm}n=1\sim n_{\max}.
\label{eq:para_Gaussian} 
\end{equation}
It is noted that the prescription is found to be very useful in optimizing
 the ranges with a small number of free parameters with high
  accuracy~\cite{Kamimura88}.
 
The total Hamiltonian for the $3\alpha$ system is presented as
\begin{eqnarray}
\mathcal{H}=\sum_{i=1}^3 T_i-T_{cm}+\sum_{i<j=1}^3\left[V_{2\alpha}(r_{ij})+V_{2\alpha}^{Coul}(r_{ij})\right]
              +V_{3\alpha}(r_{12},r_{23},r_{31})+V_{\rm Pauli},\label{H_3alpha}
\end{eqnarray}
where $T_i$, $V_{2\alpha}$ and $V_{3\alpha}$ stand for the kinetic energy operator for the {\it i}-th
 $\alpha$ particle, phenomenological $2\alpha$ and $3\alpha$ potentials, respectively,
 and $V_{2\alpha}^{Coul}$ is the Coulomb potential between 2$\alpha$ particles.
The center-of-mass kinetic energy is subtracted from the Hamiltonian.
The Pauli-blocking operator $V_{\rm Pauli}$~\cite{Kukulin84} is represented as
\begin{eqnarray}
&&V_{\rm Pauli}=\lim_{\lambda \rightarrow \infty} \lambda \hat{O}_{\rm Pauli},\label{eq_A:Pauli}\\ 
&&\hat{O}_{\rm Pauli}=\sum_{2n+\ell<4,\ell=even}
              \sum_{(ij)=(12),(23),(31)}\left|u_{n\ell}(\Vec{r}_{ij}\rangle\langle u_{n\ell}(\Vec{r}_{ij})\right|,
              \label{Pauli_operator}
\end{eqnarray}
which removes the Pauli forbidden states, $u_{00}$, $u_{10}$ and $u_{20}$, between
 any two $\alpha$ particles from the 3$\alpha$ model space.
The Gaussian size parameter of the nucleon $(0s)$ orbit in the $\alpha$ cluster
 is taken to be $b_N=1.358$ fm, which reproduces the size of the $\alpha$ particle
 in free space.
In the present study, we take the harmonic oscillator wave functions as the Pauli
 forbidden states.
The eigenenergy $E$ of $^{12}$C and coefficients $A_c$ in Eq.~(\ref{gaussian_basis}) 
 are obtained in terms of the variational principle,
\begin{eqnarray}
\delta\left[\langle\Phi_J\mid \mathcal{H}-E \mid \Phi_J\rangle\right]=0.
\end{eqnarray}

We use an effective $2\alpha$ potential which reproduces the observed $\alpha\alpha$
 scattering phase shifts ($S$-, $D$- and $G$-waves) and the resonant ground-state
 energy within the $2\alpha$ OCM framework.
The effective $2\alpha$ potential and Coulomb potential, $V_{2\alpha}$ and $V_{2\alpha}^{Coul}$, 
 are constructed with the folding procedure, where we fold the modified Hasegawa-Nagata 
 effective $NN$ interaction (MHN) and the $pp$ Coulomb potential with the $\alpha$-cluster density.
Also the strength of the second-range triplet-odd part in MHN is modified 
 so as to reproduce the $2\alpha$ scattering phase shifts.

Only using the effective $2\alpha$ potential leads to a significant overbinding energy
 for the ground state of $^{12}$C within the frame of the $3\alpha$ OCM.
Thus, we introduce an effective, repulsive, $3\alpha$ potential, $V_{3\alpha}$,
 in addition to the $2\alpha$ potential,
\begin{eqnarray}
V_{3\alpha}=V_0 \exp\left[-\beta \left(\Vec{r}_{12}^2+\Vec{r}_{23}^2+\Vec{r}_{31}^2\right)\right], 
\end{eqnarray}
where $\Vec{r}_{ij}$ denotes the relative coordinate between the {\it i}- and {\it j}-th $\alpha$
 particles, and $V_0$ and $\beta$ are taken to be $V_0=87.5$ MeV and $\beta=0.15$ fm$^{-2}$.
Including the $3\alpha$ potential, the energy of the ground state of $^{12}$C is reproduced
 with respect to the $3\alpha$ threshold, together with the nuclear radius (see Sec.~III).

Single-$\alpha$ orbits and corresponding occupation probabilities for $0^+$,
 $2^+$, $1^-$, and $3^-$ states of $^{12}$C are investigated by solving the eigenvalue equation 
 of the single $\alpha$-particle density matrix in Eqs.~(\ref{one_body_density}) 
 and (\ref{eigenvalue_equation_one_body_density}) [see Sec.~II(b)].
They will lead to a deep understanding about the structure of $^{12}$C.

In the present investigation, we make a further structure study for the $0^+$ states of $^{12}$C,
 because they have very intriguing features. 
According to Ref.~\cite{Tohsaki01},
 the $0^+_1$ state has a compact shell-model-like state, while the $0^+_2$ one 
 is conjectured to have a dilute $3\alpha$ condensate structure, the nuclear 
 radius of which is 4.3~fm, much larger than that of the ground $0^+_1$ state (2.48~fm).
Thus, it is interesting to see in detail the structure change of the $0^+$ state of $^{12}$C
 by taking the nuclear radius as a parameter.  
We investigate the dependence of the occupation probabilities and radial behaviors 
 of the single $\alpha$-particle orbits in the $0^+$ state on its rms radius
 within the $3\alpha$ OCM framework. 
The results will give us helpful understanding about the structure of $^{12}$C.
The procedure of evaluating them is formulated hereafter.

First, we consider a Pauli-principle respecting $3\alpha$ OCM basis wave function.
For the purpose, the eigenvalue problem for the Pauli operator in Eq.~(\ref{Pauli_operator}) 
 is solved to obtain the Pauli forbidden state in the $3\alpha$ OCM model space 
\begin{eqnarray}
\hat{O}_{\rm Pauli}|G_P^{3\alpha}\rangle=\lambda_P |G_P^{3\alpha}\rangle,\label{eigen_eq_Pauli_operator}
\end{eqnarray}
where $\lambda_P$ denotes the eigenvalue for the eigenfunction $|G_P^{3\alpha}\rangle$.
The Pauli operator, then, is expressed as
\begin{eqnarray}
\hat{O}_{\rm Pauli}=\sum_P {|G_P^{3\alpha}\rangle}\lambda_P{\langle G_P^{3\alpha}|}.
\end{eqnarray}
If $\lambda_P$ is non-zero, its eigenfunction corresponds to the Pauli forbidden state.
In the present study, the eigenvalue problem is solved with use of the $3\alpha$ OCM basis
 in Eq.~(\ref{3alpha_OCM_basis}).
Then, the Pauli-principle respecting OCM basis wave function is given by
\begin{eqnarray}
\tilde{\Phi}_c^{3\alpha}(\nu,\mu)=N_c(\nu,\mu)
       \left[\Phi_c^{3\alpha}(\nu,\mu)
             -\sum_{\lambda_P\not{=}0} {|G_P^{3\alpha}\rangle} 
              {{\langle G_P^{3\alpha}}|\Phi_c^{3\alpha}(\nu,\mu)\rangle}\right],
\label{pauli_free_basis}
\end{eqnarray} 
where $N_c$ denotes the normalization factor with the angular
 momentum channel $c=(\ell,L)_J$, and $\Phi_c^{3\alpha}(\nu,\mu)$ 
 is given in Eq.~(\ref{3alpha_OCM_basis}). 

According to the results of the $3\alpha$ OCM calculation (see Sec.~IIIA),
 the ground state ($0^+_1$) and second $0^+_2$ states of $^{12}$C
 have the equilateral triangle configuration of the $3\alpha$ clusters.
In addition, it is found that the single-angular-momentum-channel calculation
 with $c=(\ell L)_J=(00)_0$ gives a good approximation to the results
 of the full coupled-channel calculation. 
Thus, only the single angular momentum channel $c=(\ell L)_J=(00)_0$
 is taken in the present calculation, and the equilateral triangle 
 configuration is assumed for the basis wave function.
The latter can be realized easily by putting the condition $\nu=\mu$ in Eq.~(\ref{pauli_free_basis}).
Then, the rms radius of the Pauli-principle respecting $3\alpha$ OCM basis wave function is evaluated as
\begin{eqnarray}
\sqrt{\langle r^2\rangle_{\nu}}=\left[  \left\langle \tilde{\Phi}_{(00)_0}^{3\alpha}(\nu,\mu=\nu) |
                        \sum_{i=1}^3 {\Vec{r}_i^{(cm)}}^2
                        |\tilde{\Phi}_{(00)_0}^{3\alpha}(\nu,\mu=\nu) \right\rangle + 1.71^2 \right]^{1/2},
\end{eqnarray}
which depends on $\nu$, and where we take into account the finite size of the $\alpha$ cluster.
The energy of the $3\alpha$ system is given by
\begin{eqnarray}
E_{3\alpha}(\nu)=\left\langle \tilde{\Phi}_{(00)_0}^{3\alpha}(\nu,\mu=\nu) |
                        \tilde{\mathcal H}
                        |\tilde{\Phi}_{(00)_0}^{3\alpha}(\nu,\mu=\nu) \right\rangle.
\end{eqnarray}
where $\tilde{\mathcal H}$ denotes the total Hamiltonian of the $3\alpha$ system in which
 we subtract the Pauli-blocking operator $V_{\rm Pauli}$ from the $3\alpha$ OCM Hamiltonian
 ${\mathcal H}$ in Eq.~(\ref{H_3alpha}).  
The single-$\alpha$ density matrix is given as
\begin{eqnarray}
\rho_{\nu}(\Vec{r},\Vec{r}')=\langle \tilde{\Phi}_{(00)_0}^{3\alpha}(\nu,\mu=\nu) \mid 
 {\sum_{i=1}^3 \mid \delta(\Vec{r_i^{(cm)}}-\Vec{r}')\rangle\langle\delta(\Vec{r_i^{(cm)}}-\Vec{r})} 
 \mid \tilde{\Phi}_{(00)_0}^{3\alpha}(\nu,\mu=\nu) \rangle.
\end{eqnarray}
The single $\alpha$-particle orbit and its occupation number in the basis wave function
 are obtained by solving the eigenvalue equation of the single-$\alpha$ density matrix,
\begin{eqnarray}
\int d\Vec{r}'\rho_{\nu}(\Vec{r},\Vec{r}')\varphi_\eta(\Vec{r}')=\eta \varphi_\eta(\Vec{r}),\label{eigen_eq_density_rms}
\end{eqnarray}
where the eigenfunction $\varphi_\eta$ denotes the $\alpha$-particle orbit 
 with the occupation number $\eta$ (eigenvalue).
Thus, we can study the dependence of  the occupancy of the single $\alpha$-particle orbits
 in the $0^+$ state of $^{12}$C on its nuclear radius by choosing the parameter value $\nu$
 so as to reproduce a given nuclear radius.

\newpage
\section{Results and discussion}

\subsection{$0^+_1$ and $0^+_2$ states}

\begin{figure}
\begin{center}
\includegraphics*[scale=0.5]{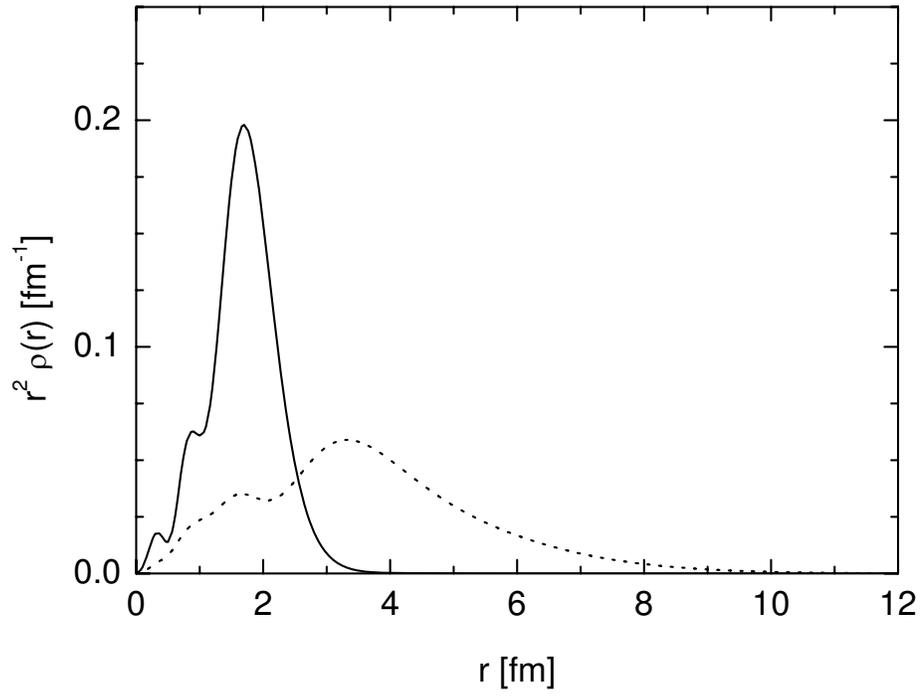}
\caption{
Density distribution of the $\alpha$ particle for the $0^+_1$ (solid line)
 and $0^+_2$ (dotted) states.
}
\label{fig:2}
\end{center}
\end{figure}

Table~\ref{tab:1} presents the results for the energy, measured from the $3\alpha$ 
 threshold, and nuclear radii for the ground ($0^+_1$) and excited states
 ($0^+_2$) of $^{12}$C.
The energy for the ground state is reproduced well, and  
 the corresponding nuclear radius, 2.44 fm, is in good agreement with the experimental
 charge radius ($2.4829\pm0.019$ fm) with an error of about $2\%$. 
On the other hand, the rms distance between $2\alpha$ clusters in the $0^+_1$ state is 
 $\sqrt{\langle r^2\rangle_{\alpha\alpha}}$=3.02 fm (see Table~\ref{tab:1}), 
 which is larger than that between the center-of-mass of the $2\alpha$ clusters 
 and the third $\alpha$ cluster, $\sqrt{\langle r^2\rangle_{\alpha-2\alpha}}$=2.61 fm.
Then, the square of the ratio,
 $\left[\sqrt{\langle r^2\rangle_{\alpha-2\alpha}}/\sqrt{\langle r^2\rangle_{\alpha\alpha}}\right]^2$,
 is about 3/4. 
The results mean that the ground state has an equilateral-triangle-like intrinsic shape.
Figure~\ref{fig:2} shows the density distribution of the $\alpha$ particle for the $0^+_1$
 state of $^{12}$C.
We see a prominent peak at $r\sim 2$ fm, which demonstrates clearly 
 the shell-model-like compact structure of the ground state of $^{12}$C. 

As for the $0^+_2$ state, the energy measured from the $3\alpha$ threshold 
 is $E_{3\alpha}$=0.86 MeV ($E_x$=8.13 MeV), 
 which agrees well with the experimental data $E^{exp}_{3\alpha}$=0.38 MeV 
 ($E^{exp}_x$=7.65 MeV).
The calculated nuclear radius for the $0^+_2$ state is as large as 4.31 fm (see Table~\ref{tab:1}).
This means that the state has a dilute $3\alpha$ structure, although our nuclear radius 
 is a little larger than that in Ref.~\cite{Tohsaki01}.
The density distribution of the $\alpha$ particle for the $0^+_2$ state is illustrated
 in Fig.~\ref{fig:2}.
In comparison with that for the ground state, we can easily recognize the dilute structure 
 of the $0^+_2$ state, which is in contrast with the compact structure of the ground state.
 
\begin{figure}
\begin{center}
\includegraphics*[scale=0.5,clip]{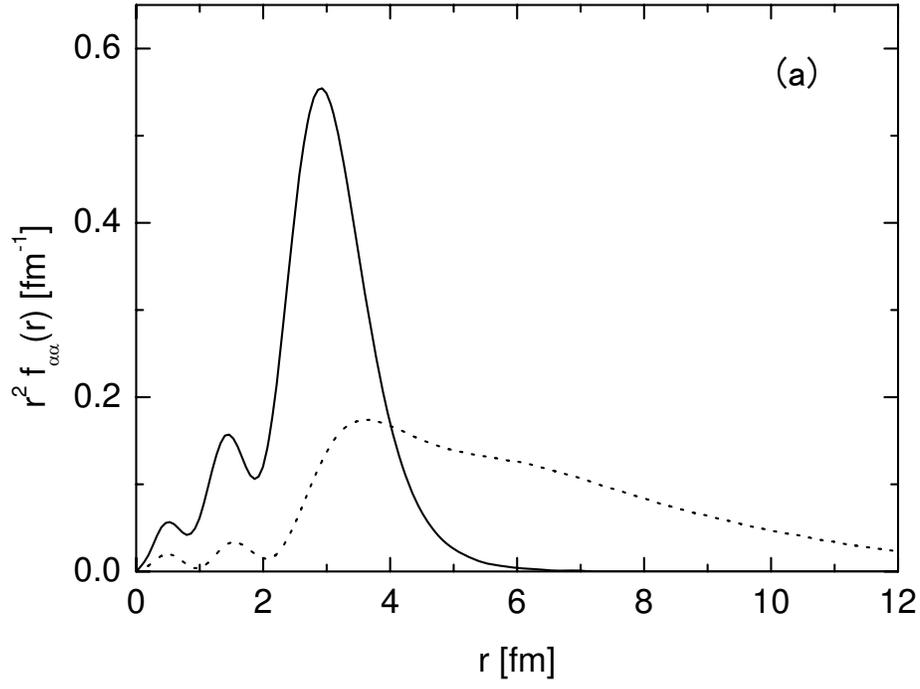}
\includegraphics*[scale=0.5,clip]{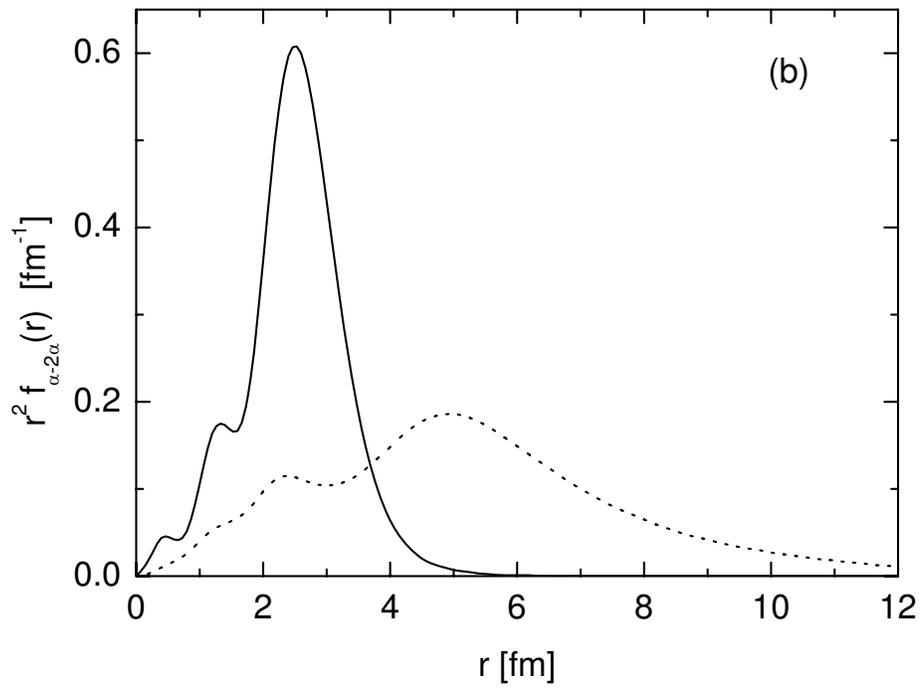}
\caption{
Correlation functions, (a)~$f_{\alpha\alpha}$ and (b)~$f_{\alpha-2\alpha}$,
 for the $0^+_1$ (solid line) and $0^+_2$ (dotted) states. 
}
\label{fig:3}
\end{center}
\end{figure}

The difference between the structures of the $0^+_1$ and $0^+_2$ states
 can be also seen in the radial behavior of the correlation functions,
 $f_{\alpha\alpha}$ and $f_{\alpha-2\alpha}$,
 with respect to the $\alpha$-$\alpha$ and $\alpha$-$2\alpha$ relative
 coordinates, respectively, of Eqs.~(\ref{cf_2alpha}) and (\ref{cf_alpha_2alpha}).
They are illustrated in Fig.~\ref{fig:3}.
Reflecting the compact structure of the $0^+_1$ state, both
 of $f_{\alpha\alpha}$ and $f_{\alpha-2\alpha}$ have prominent peaks 
 at $r\sim2.6$ fm and $2.5$ fm, respectively, and extend to $r\sim5$ fm,
 while those for the $0^+_2$ state show bump structures with
 peaks at $r\sim$4 fm and $r\sim$5 fm, respectively, 
 and have a long tail up to $r\sim15$ fm.
 
It is instructive to study the single $\alpha$-particle orbits (eigenfunctions) 
 and occupation numbers (eigenvalues) of the one-body density 
 matrix in Eq.~(\ref{one_body_density}).
The results of the diagonalization of the latter are shown in Table~\ref{tab:2} together with the occupation
 probability defined as the occupation number divided by the number of $\alpha$ particles.
As for the ground state, the occupation probabilities spread out over
 $S$, $D$ and $G$ waves, but they are concentrated to the first
 orbits, $S_1$, $D_1$ and $G_1$ orbits, respectively, where
 $L_k$ denotes the $k$-th orbit for the $L$ wave.
The occupation probabilities are about $30~\%$ for all orbits.
This result is expected from the fact that the ground-state wave function is
 of nuclear SU(3)-like character, SU(3)$[f](\lambda\mu)_{J^\pi}=[444](04)_{0^+}$
 with quanta $Q$=8, where the SU(3) bases with $Q<8$ correspond to the Pauli-forbidden states. 
Since the SU(3)$[444](04)_{0^+}$ state is the eigenfunction of the $3\alpha$ RGM norm kernel
 in Eq.~(\ref{eigenvalue_eq}), it can be regarded as the $3\alpha$ boson wave function
 with $Q=8$, see Eq.~(\ref{boson_wf_2}). 
The state is described as 
\begin{eqnarray}
\left|[444](04)\right\rangle_{0^+}&=&\sum_{n\ell N L} a_{nlNL}{|{(n\ell)}{(NL)}\rangle},\nonumber\\
      &=& \sqrt{\frac{64}{225}}|2s2S\rangle
      -\sqrt{\frac{80}{225}}|1d1D)\rangle
      +\sqrt{\frac{81}{225}}|0g0G\rangle,\label{SU3_0+}
\end{eqnarray}
 where $|(n\ell)(NL)\rangle$ presents the basis function 
 $|u_{n\ell}(\Vec{r}_{2\alpha})u_{NL}(\Vec{r}_{\alpha-2\alpha})\rangle$
 with $2n+\ell+2N+L=8$, and $u_{n\ell}$ ($u_{NL}$) denotes the harmonic oscillator wave function with
 the number of nodes $n$ ($N$) and orbital angular momentum $\ell$ ($L$)
 referring to the coordinate vector $\Vec{r}_{2\alpha}$ ($\Vec{r}_{\alpha-2\alpha}$)
 between $2\alpha$ clusters (between the center-of-mass for the $2\alpha$ 
 clusters and the third $\alpha$ cluster).
Let us define $L_\alpha$ as the orbital angular momentum of a single-$\alpha$ orbit.
Then, $L$ in Eq.~(\ref{SU3_0+}) corresponds to $L_\alpha$,  
 because $L_\alpha$ is defined as the orbital angular momentum with respect to $\Vec{r}_\alpha^{(cm)}$, 
 coordinate vector of the $\alpha$ particle measured from the center-of-mass coordinate 
 of $^{12}$C (see Eq.~(\ref{single_density})), which is parallel to $\Vec{r}_{\alpha-2\alpha}$ 
 (${\Vec{r}_\alpha^{(cm)}}=\frac{2}{3}\Vec{r}_{\alpha-2\alpha}$).
From the definition of the one-body density matrix in Eq.~(\ref{one_body_density}), 
 the single-$\alpha$ orbits and occupation probabilities for the SU(3) state in Eq.~(\ref{SU3_0+})
 are given as follows:~$64/255\sim28 \%$ for $S$-orbit, $80/225\sim36 \%$ for $D$-orbit,
 and $81/225\sim36 \%$ for $G$-orbit.
Thus, we can understand the reason why the $S_1$, $D_1$ and $G_1$
 orbits in Table~\ref{tab:2} have about $30 \%$ occupation probabilities each.
 
\begin{figure}
\begin{center}
\includegraphics*[scale=0.5,clip]{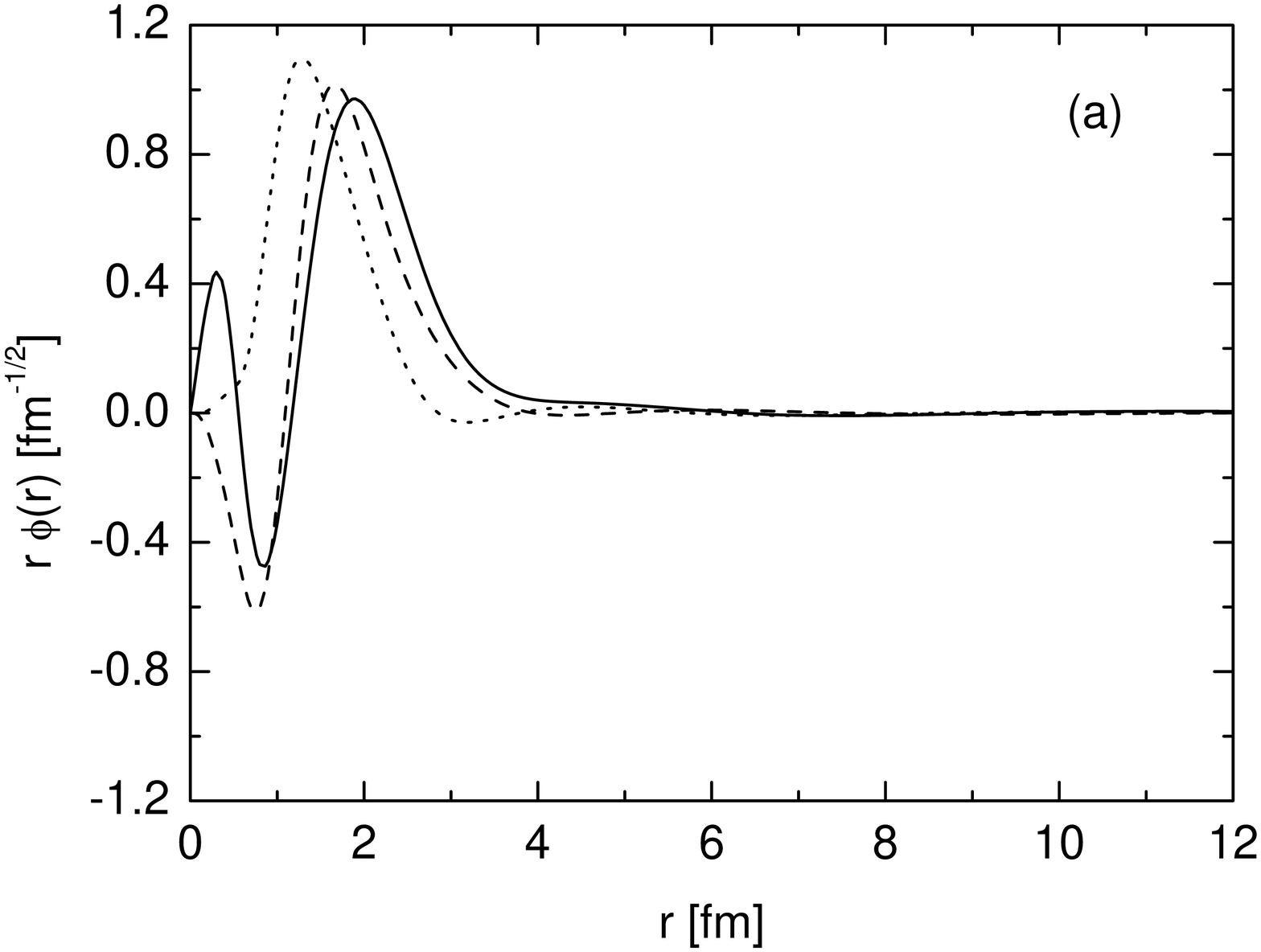}
\includegraphics*[scale=0.5,clip]{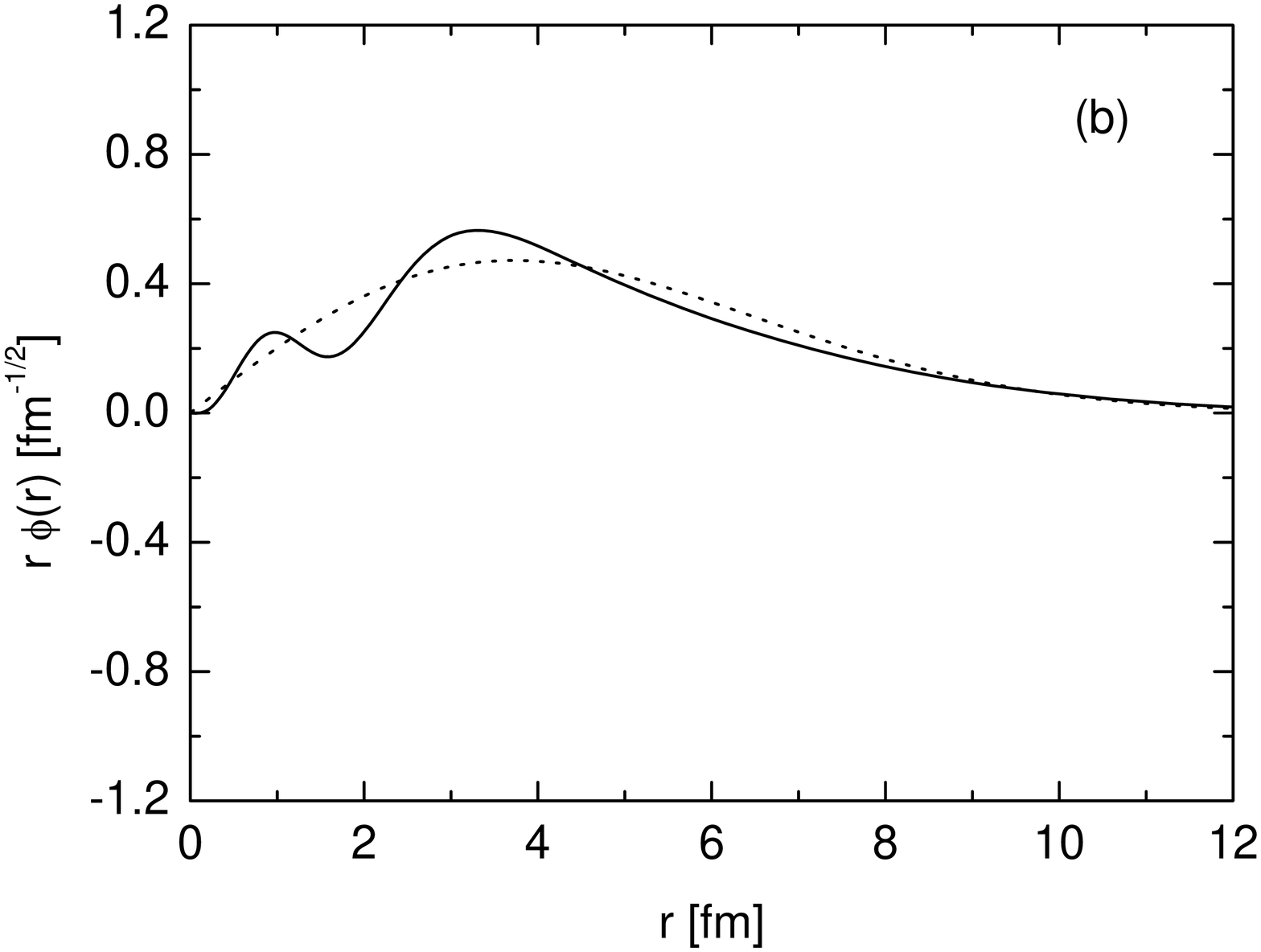}
\caption{
Radial parts of the single $\alpha$ orbits, (a)~$S_1$ (solid line), $D_1$ (dashed) 
 and $G_1$ (dotted), in the $0^+_1$ state, 
 and (b)~the $S_1$ (solid) orbit in the $0^+_2$ state and $S$-wave
 Gaussian function (dotted), $r\varphi_{0s}$, with the size parameter 
 $a=0.038$ fm$^{-2}$ (see text).
Note that all of the radial parts in figures are multiplied by $r$.
}
\label{fig:4}
\end{center}
\end{figure}

\begin{figure}
\begin{center}
\includegraphics*[scale=0.5,clip]{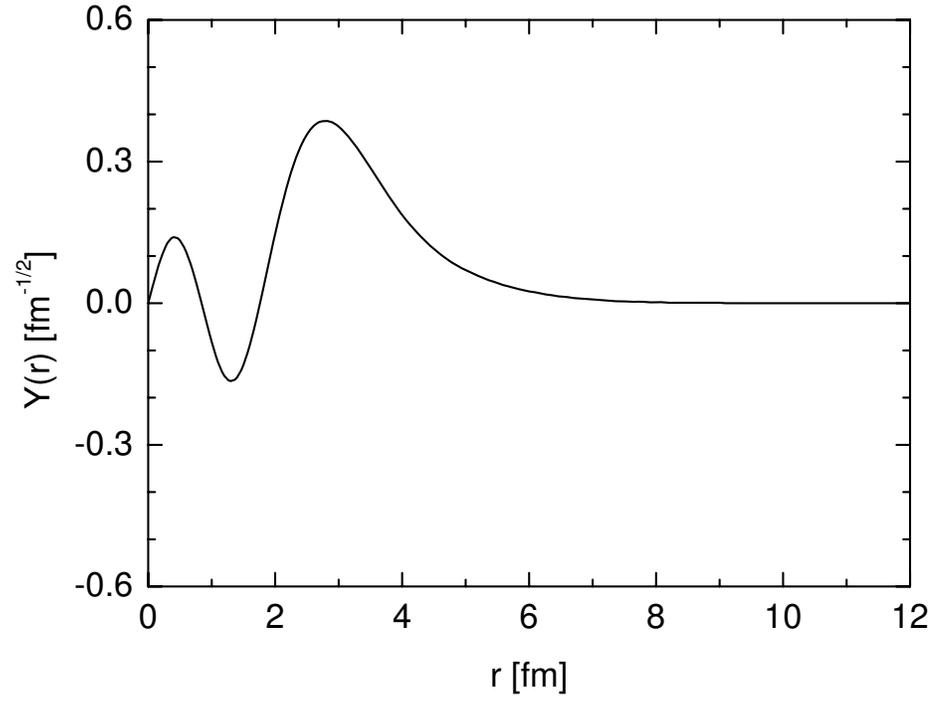}
\caption{
Reduced width amplitude of the $\alpha$+$^8$Be($0^+$) channel
 for the $0^+_1$ state. 
}
\label{fig:5}
\end{center}
\end{figure}

Figure~\ref{fig:4}(a) demonstrates the radial parts for the $S_1$-, $D_1$- 
 and $G_1$-orbits, the number of nodes of which are two, one and zero, respectively.
Reflecting the SU(3) character, the radial behaviors of the three orbits are similar 
 to those of the harmonic oscillator wave functions ($u_{NL}$) with $Q=4$, $u_{02}$, 
 $u_{21}$ and $u_{40}$, respectively, where $N$ ($L$) denotes 
 the number of nodes (orbital angular momentum).
We see that the radial parts of the single $\alpha$-particle orbits oscillate widely 
 in the inside region ($r<4$ fm). 
This is due to the strong Pauli blocking effect for the ground state with
 the compact shell-model-like structure.
The large oscillation can also be seen in the reduced width amplitude
 of the $\alpha$+$^8$Be($0^+$) channel for the ground state 
 shown in Fig.~\ref{fig:5}.

Concerning the $0^+_2$ state, the occupation probabilities are shown in Table~\ref{tab:2}. 
We see a strong concentration on a single orbit:~the occupation probability of the
 $S_1$ orbit is largest, amounting to about $70 \%$, and those for other orbits are very small.
This means that each of the three $\alpha$ particles in the $0^+_2$ state is 
 in the $S_1$ orbit with occupation probability as large as about $70 \%$.
The radial behavior of the $S_1$ orbit is illustrated with the solid line in Fig.~\ref{fig:4}(b).
We see no nodal behavior but small oscillations in the inner region ($r<4$ fm) 
 and a long tail up to $r\sim$10 fm.
For reference, the radial behavior of the $S$-wave Gaussian
 function, $\varphi_{0s}(r)=N_{0s}(a)\exp(-ar^2)$, is drawn with the dashed line
 in Fig.~\ref{fig:4}(b), where the size parameter $a$ is chosen to be 
 0.038 fm$^{-2}$, and $N_{0s}(a)$ denotes the normalization factor.
The radial behavior of the $S_1$ orbit is similar
 to that of the $S$-wave Gaussian function, in particular, in the outer region ($r>4$ fm),
 whereas a slight oscillation of the former around the latter can be seen 
 in the inner region ($r<4$ fm).

\begin{figure}
\begin{center}
\includegraphics*[scale=0.5,clip]{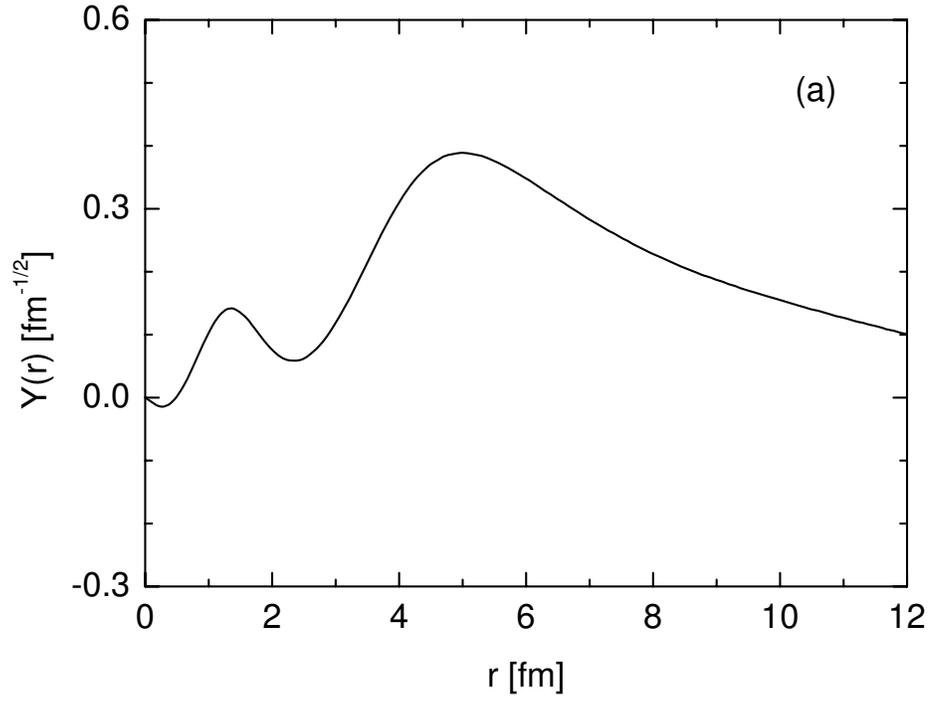}
\includegraphics*[scale=0.5,clip]{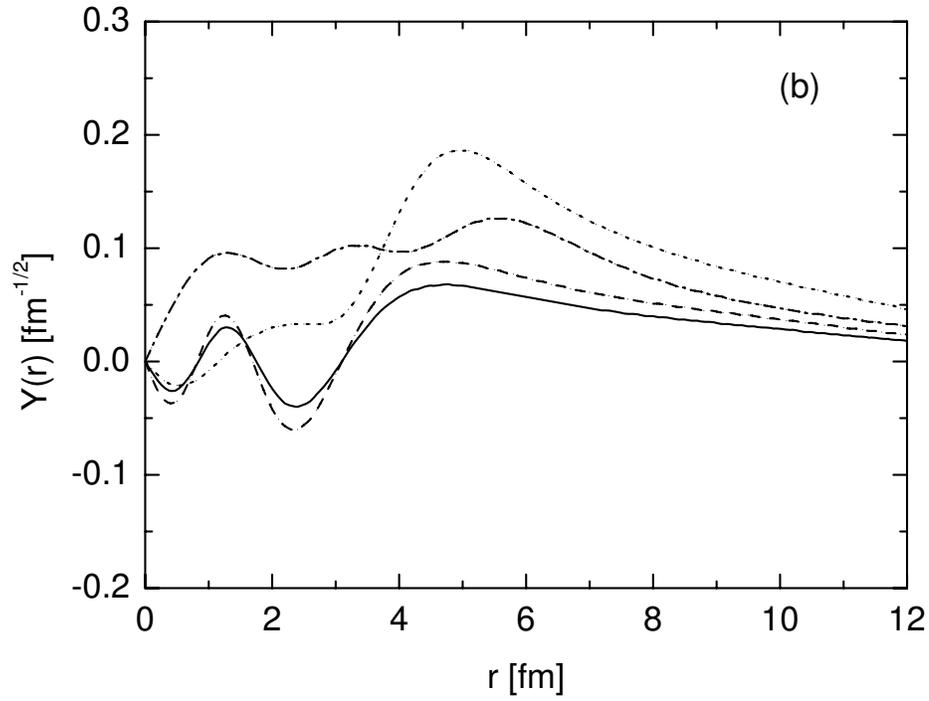}
\caption{
(a)~Reduced width amplitude of the $\alpha$+$^8$Be($0^+$) channel
 for the $0^+_2$ state, and
(b)~those in which the distance between
 the $2\alpha$ clusters in $^8$Be is fixed to 
 $r_{\alpha\alpha}$=0.5, 2.5, 4.5 and 6.5 fm. 
}
\label{fig:6}
\end{center}
\end{figure}

The small oscillation of the $S_1$ orbit in the inner region can also be seen 
 in the reduced width amplitude of the $0^+_2$ state for the $\alpha$-$^8$Be($0^+$) 
 channel in Fig.~\ref{fig:6}(a). 
In order to study the origin of the small oscillation,
 we show in Fig.~\ref{fig:6}(b) the results of the reduced width amplitudes of the $0^+_2$ state 
 for the $\alpha$-$^8$Be($0^+$) channel, fixing the distance between
 the $2\alpha$ clusters in $^8$Be to $r_{\alpha\alpha}$=0.5, 2.5, 4.5 and 6.5 fm.  
In the case of $r_{\alpha\alpha}< 4$ fm, we see the nodal behavior with the large
 oscillation in the inner region, coming from the strong Pauli blocking effect
 among the $3\alpha$ clusters, while the nodal behavior is disappearing and
 the oscillations are getting weaker for the larger $r_{\alpha\alpha}$ ($\geq 4$ fm),
 reflecting the weaker Pauli blocking effect.
Thus, the small oscillations in the radial behavior of the $S_1$ orbit
 is evidence for the weak Pauli blocking effect for the $0^+_2$ state with the dilute structure.

\begin{figure}
\begin{center}
\includegraphics*[scale=0.5,clip]{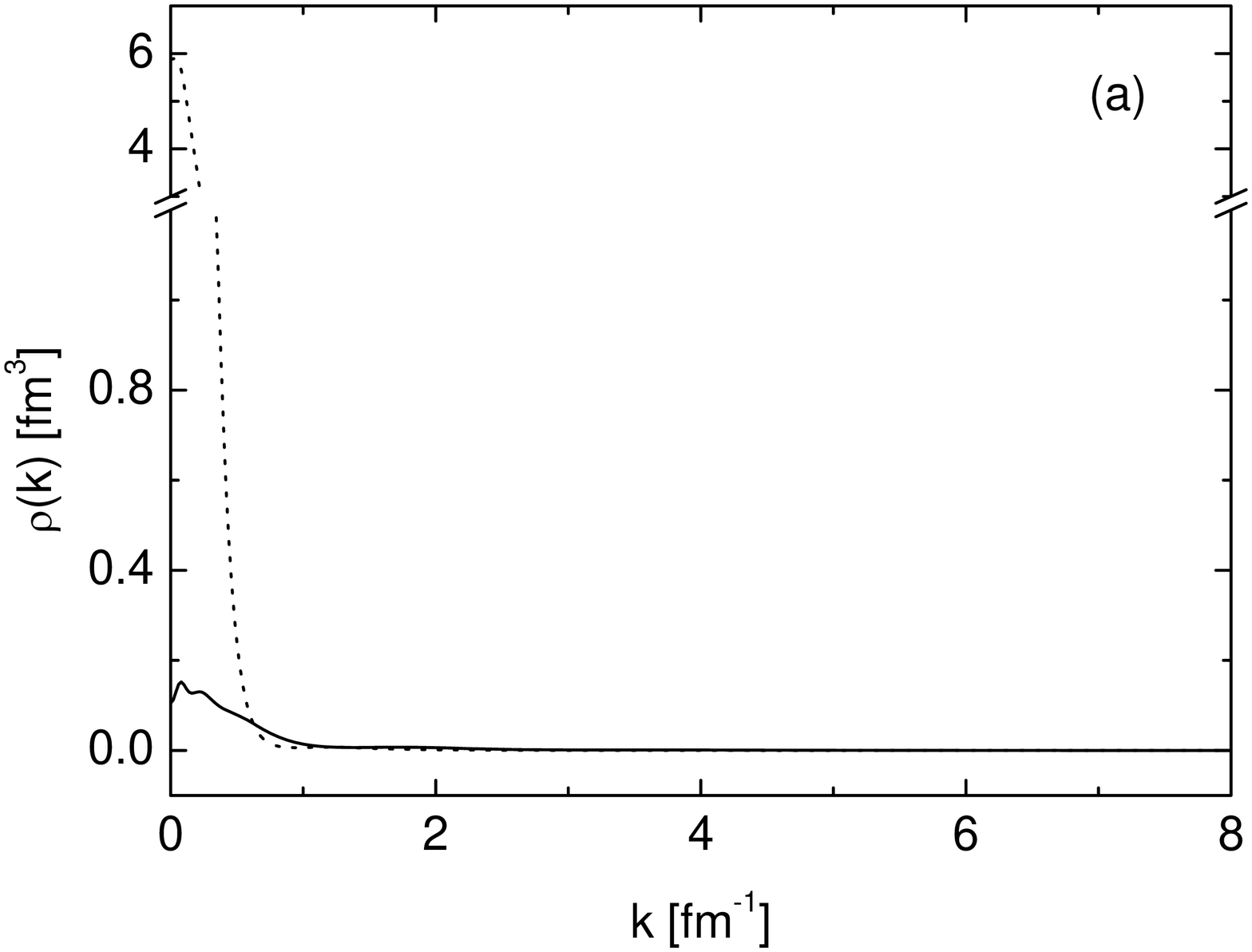}
\includegraphics*[scale=0.5,clip]{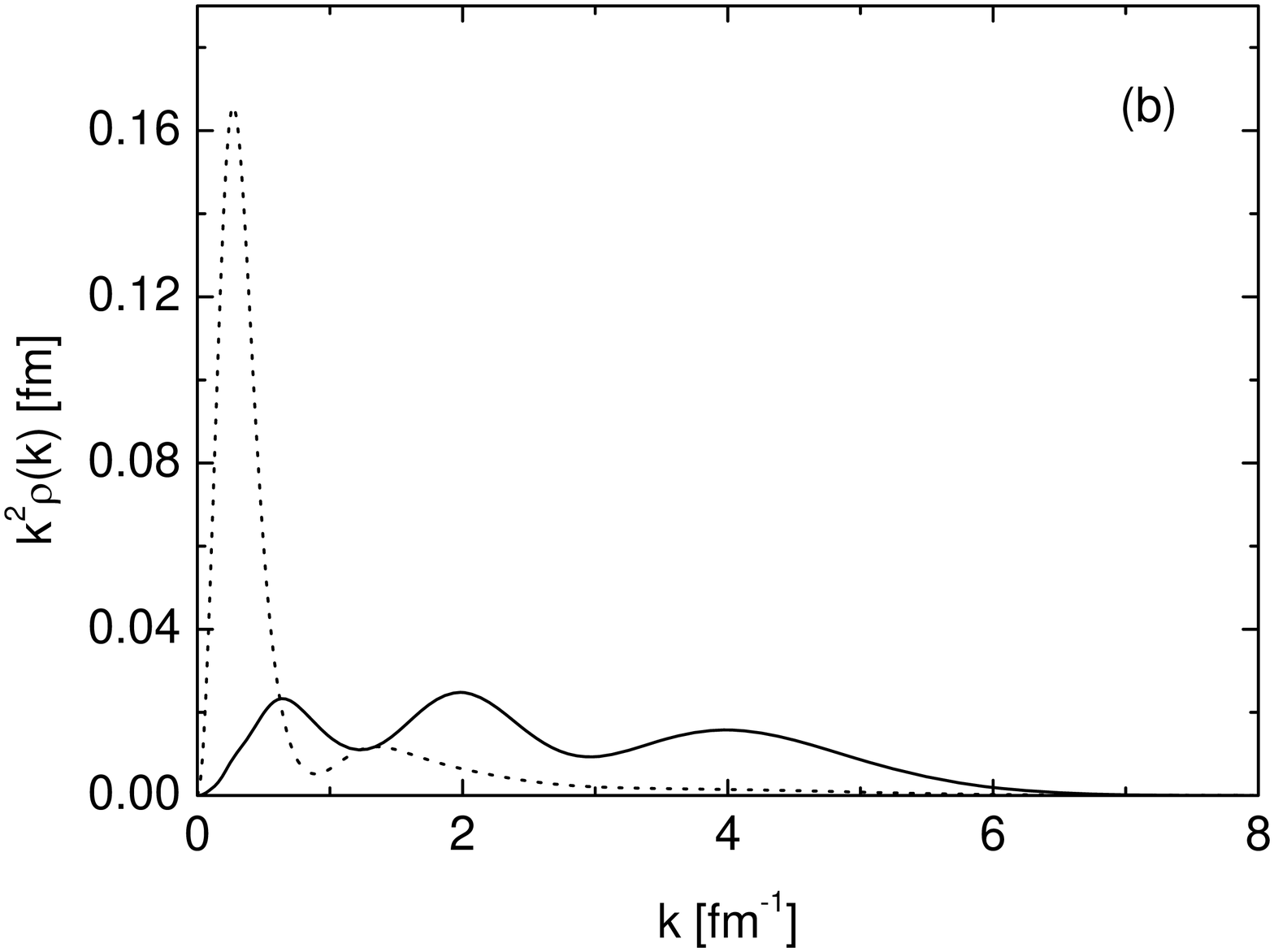}
\caption{
Momentum distribution of the $\alpha$ particle, (a)~$\rho(k)$ and (b)~$k^2\rho(k)$,
 for the $0^+_1$ (solid line) and $0^+_2$ (dotted) states. 
}
\label{fig:7}
\end{center}
\end{figure}

The momentum distributions of the $\alpha$ particles, $\rho(k)$ and $k^2\times\rho(k)$, 
 are shown for the $0^+_1$ and $0^+_2$ states in Fig.~\ref{fig:7}.
Reflecting the dilute structure of the $0^+_2$ state, we see a strong concentration 
 of the momentum distribution in the $k<1$ fm$^{-1}$ region, 
 and the behavior of $\rho(k)$ is of the $\delta$-function type, similar to 
 the momentum distribution of the dilute neutral atomic condensate states
 at very low temperature trapped by the external magnetic field~\cite{Dalfovo99}.
On the other hand, the ground state has higher momentum component 
 up to $k\sim6$ fm$^{-1}$ as seen from the behavior of $k^2\times\rho(k)$ reflecting 
 the compact structure.
The above results for the radial behavior of the $S_1$ orbit,
 occupation probability and momentum distribution for the $0^+_2$ state leads us 
 to conclude that this state is similar to an ideal dilute $3\alpha$ condensate
 with as much as about $70 \%$ occupation probability.

Let us make some remarks on the calculated energy ($E_{3\alpha}$=0.85 MeV) 
 and wave function of the $0^+_2$ state.
They were evaluated under the bound state approximation in the present study (see Sec.~II).
The quite small experimental width for $0^+_2$ ($\Gamma$=8.5 eV)~\cite{Ajzenberg90} means 
 that the bound state approximation is very good to describe the wave function.
The complex scaling method~\cite{Kuruppa88} is powerful to search for resonant states 
 and calculate the exact energies and widths, and is applied easily to the $3\alpha$ system 
 by slightly modifying the present framework. 
The detailed framework is skipped here and referred to Ref.~\cite{Kuruppa88}.
In the present study, we investigated the energy of the $0^+_2$ state with the complex scaling method.
It was found that a resonant state, corresponding to the $0^+_2$ state, appears 
 at $E_{3\alpha}$=0.85 MeV with a width less than the numerical uncertainty
 ($\sim$100 keV in the present calculation).
The results confirm that the bound state approximation is good to describe the $0^+_2$ resonant state.

It is interesting to compare our results with those given by
 Matsuura et al.~\cite{Matsuura04}, who used the normalized spectroscopic amplitude 
 to obtain the bosonic quantities such as the single-$\alpha$ orbits
 and occupation probabilities for the $0^+_2$ state
 in place of the $3\alpha$ boson wave function.
According to their results, the occupation probability of the $S_1$ orbit
 ($0S$ orbit in Ref.~\cite{Matsuura04}) for the $0^+_2$ state is about $70 \%$,
 the value of which is quite similar to ours in Table~\ref{tab:2}.
However, the radial behavior of the $S_1$ orbit for the $0^+_2$ state
 as well as the one of the $0^+_1$ state given by Matsuura et al.~are 
 quite different from our results, and seems unnatural.
For example, the $S_1$ orbit for the $0^+_2$ state
 has as much as $6\sim8$ nodes and shows a behavior similar to that for the $0^+_1$ state, 
 in spite of the fact that the $0^+_2$ state has a dilute $3\alpha$ condensate 
 structure (see Fig.~6 in Ref.~\cite{Matsuura04}). 
In addition, the $G$-orbit for $0^+_1$ state has a prominent
 peak at $r\sim13$ fm, although the state has a shell-model-like compact structure.
Also the radial behavior of the single-$\alpha$ orbits given in Ref.~\cite{Matsuura04} 
 is hard to understand.
This may be due to the fact that those authors used the normalized spectroscopic amplitude 
 in place of the $3\alpha$ boson wave function.     

\subsection{$2^+_1$ and $2^+_2$ states}

The $2^+_1$ state at $E^{exp}_{3\alpha}=-2.83$ MeV ($E_x$=4.44 MeV) belongs to 
 the rotational band of the ground state starting from the $0^+_1$ state at $E^{exp}_{3\alpha}=-7.27$ MeV.
The calculated energy and nuclear radius for $2^+_1$ in the present study 
 are shown in Table~\ref{tab:1}:~$E_{3\alpha}=-5.28$ MeV and 2.45 fm, respectively.
The nuclear radius is almost the same as the one for the ground state, although 
 the calculated excitation energy is underestimated in comparison with the experimental one,
 in line with what is discussed in other papers with the microscopic or semi-microscopic $3\alpha$ 
 cluster model~\cite{Fujiwara80,Saito68,Horiuchi74,Kato89,Kurokawa04}.

\begin{figure}
\begin{center}
\includegraphics*[scale=0.5,clip]{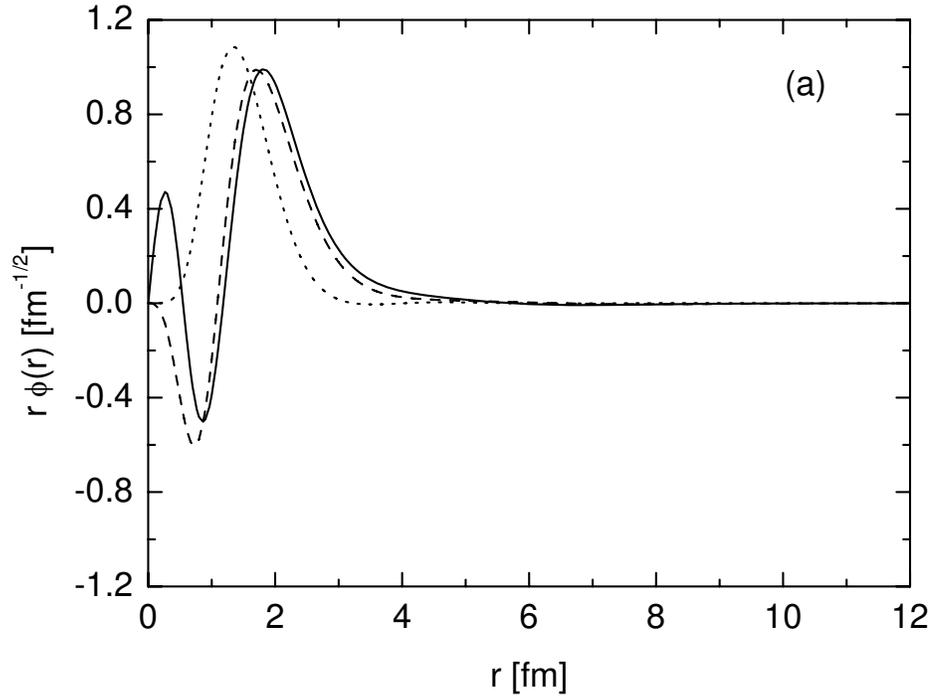}
\includegraphics*[scale=0.5,clip]{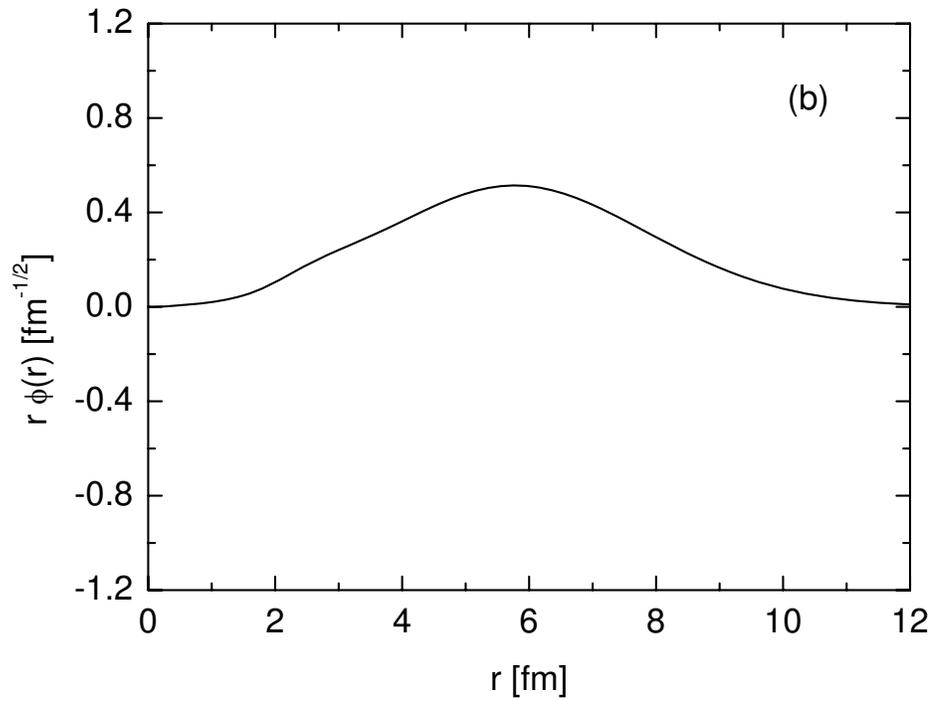}
\caption{
Radial parts of the single $\alpha$ orbits, (a)~$S_1$ (solid line), $D_1$ (dashed) 
 and $G_1$ (dotted), in the $2^+_1$ state, 
 and (b)~the $D_1$ (solid) orbit in the $2^+_2$ state.
Note that all of the radial parts in figures are multiplied by $r$.
}
\label{fig:8}
\end{center}
\end{figure}

The occupation probabilities of the single-$\alpha$ orbits for $2^+_1$ are demonstrated 
 in Table~\ref{tab:2}.
The occupation numbers are concentrated to the first $D_1$ and $G_1$ orbits with about $50 \%$.
Comparing with those for the $0^+_1$ state, we notice the smallness of the occupation number 
 for the $S_1$ orbit.
This feature can be understood from the fact that the $2^+_1$ state
 is of the SU(3)$[f](\lambda\mu)_J=[444](04)_{2^+}$ type with $Q$=8.
The SU(3) state is described as
\begin{eqnarray}
{|[444](04)\rangle_{2^+}}&=&{\sqrt{0.07111}|1d2S\rangle},\nonumber\\
                            &+&{\sqrt{0.07111}|2s1D\rangle}-{\sqrt{0.43900}|1d1D\rangle}-{\sqrt{0.00735}|0g1D\rangle},\nonumber\\
                            &-&{\sqrt{0.00735}|1d0G\rangle}+{\sqrt{0.40408}|0g0G\rangle},\label{SU3_2+}                           
\end{eqnarray}
 where $|(n\ell)(NL)\rangle$ presents the basis function, $|u_{n\ell}(\Vec{r}_{2\alpha})u_{NL}(\Vec{r}_{\alpha-2\alpha})\rangle$
 (with $2n+\ell+2N+L=8$), with the harmonic oscillator wave function $u_{n\ell}$.
From the definition of the one-body density matrix in Eq.~(\ref{one_body_density}), 
 the occupation probabilities for the SU(3) state in Eq.~(\ref{SU3_2+}) are given 
 as 0.07111 for $S$ orbit, 0.07111+0.43900+0.00735=0.51746 for $D$ orbit, 
 and 0.00735+0.40408=0.41143 for $G$ orbit.
Reflecting the character of the SU(3) structure, the occupation probability for the $S_1$ orbit 
 in Table~\ref{tab:2} is as small as $8.5~\%$. 
The radial behavior of the single-$\alpha$ orbits, $S_1$, $D_1$ and $G_1$ ones, 
 is shown in Fig.~\ref{fig:8}(a).
They are similar to those for the $0^+_1$ state shown in Fig.~\ref{fig:4}(a).

The structure study of $^{12}$C based on the $3\alpha$-condensate type wave function~\cite{Funaki04}
 indicated that the $2^+_2$ state at $E^{exp}_{3\alpha}=2.6\pm0.3$ MeV 
 with the width of $\Gamma=1.0\pm0.3$ MeV~\cite{Itoh04}
 has a structure similar to the $0^+_2$ state at $E_{3\alpha}=0.38$ MeV with the dilute $3\alpha$
 condensation~\cite{Funaki04}.
The conclusion stems from the result that the $2^+_2$ state has a large overlap
 with the single condensate wave function of a $3\alpha$ gas-like structure, the squared value of which
 amounts to about $88 \%$.
Thus, it is interesting to study the structure of the $2^+_2$ state in the present framework.
Since the $2^+_2$ state is a resonant state with non negligible width, a continuum treatment
 is requested to estimate exactly the resonant energy and width. 

In order to study the resonant properties of the $2^+_2$ state, we take 
 the complex-scaling method~\cite{Kuruppa88}, which can be applied easily to the present $3\alpha$ system 
 by slightly modifying the framework given in Sec.~II.
The method is powerful to evaluate not only the resonant energy and width but also the nuclear radius.  
The details are again skipped here and we refer to Ref.~\cite{Kuruppa88}.
The calculated results are as follows:~1)~the $2^+_2$ resonant state is located at $E_{3\alpha}$=2.3 MeV
 with $\Gamma$=1.0 MeV, results which are in good agreement with the
 experimental data~\cite{Itoh04}, and 2)~the calculated nuclear radius is 4.3 fm, almost the same
 as that of the $0^+_2$ state. 
Thus, the $2^+_2$ state has a dilute $3\alpha$ structure.

It is interesting to study the single-$\alpha$ orbits and occupation 
 probabilities in the $2^+_2$ state.
For this purpose, we need to have the wave function of the $2^+_2$ state.
Since the calculated width is not so large in comparison with the resonance energy, 
 the bound state approximation is rather good to describe the resonant wave function.
The bound state approximation of the wave function is obtained within the framework of Sec.~II,
 although the wave function gives a large nuclear radius, about 6 fm 
 (see Table~\ref{tab:1}).
Table~\ref{tab:2} illustrates the occupation probabilities of the single-$\alpha$ orbits
 ($S$-, $D$- and $G$-waves) for the $2^+_2$ state.
We see that the occupation probability concentrates on only one orbit, the $D_1$ orbit, 
 with occupancy as large as $83~\%$, and the radial behavior of the orbit is likely
 to be of the $D$-wave Gaussian-function-type with a long tail [see Fig.~\ref{fig:8}(b)], 
 reflecting a dilute structure.
These characteristics are quite similar to those for the $0^+_2$ state.
Thus, we conclude that the $2^+_2$ state belongs to the $3\alpha$-condensate structure.

According to the results in Ref.~\cite{Funaki04}, it was found that the $2^+_2$ state 
 has dominant $S$-wave between $2\alpha$ particles and a $D$-wave 
 between the center-of-mass of the $2\alpha$ particles and the third $\alpha$,
\begin{eqnarray} 
 \Phi(2^+_2)\sim|u_{\ell=0}(\Vec{r}_{2\alpha})U_{L=2}(\Vec{r}_{\alpha-2\alpha})\rangle.\label{2nd_2+_Funaki}
\end{eqnarray}
This interpretation is consistent with the preset result. 
The reason is as the follows.
From the definition of the single-$\alpha$ density matrix in Eq.~(\ref{one_body_density}),
 the single-$\alpha$ density of the $2^+_2$ state is presented as 
\begin{eqnarray}
 \rho(\Vec{r},\Vec{r}')&=&3\times \langle\Phi(2^+_2)\mid \delta(\Vec{r_1^{(cm)}}-\Vec{r}')\rangle\langle\delta(\Vec{r_1^{(cm)}}-\Vec{r}) \mid\Phi(2^+_2)\rangle,\\
 &\sim& 3\times N_{2\alpha} \times {U_{L=2}(\Vec{r})} {U_{L=2}}^*(\Vec{r}'),
\end{eqnarray}
where $N_{2\alpha}=\int d\Vec{r}_{2\alpha} {u_0}^*(\Vec{r}_{2\alpha}) u_0(\Vec{r}_{2\alpha})\sim 1$.
Thus, the $2^+_2$-state wave function, Eq.~(\ref{2nd_2+_Funaki}), has
 a dominant occupation probability of the $D$-orbit, $U_{L=2}$.
The results are in good agreement with the present study.

\subsection{$3^-_1$ state}

The $3^-$ state at $E^{exp}_{3\alpha}=$2.37 MeV is an interesting one
 from the point of view of the dilute $\alpha$ condensation.
If the state is a condensate with all of the $3\alpha$
 particles in the $P$ orbit, there is the possibility of a superfuid with vortex lines,
 similar to the rotating dilute atomic condensate at very low temperature~\cite{Dalfovo99}.
Thus, it is an intriguing problem to study the structure in the present framework.

The calculated energy of the $3^-$ state is in good agreement with the experimental 
 data (see Table~\ref{tab:1}).
The very small width ($\Gamma^{exp}=3.4$ keV)~\cite{Ajzenberg90} indicates that the bound state
 approximation is very good to describe the state.
In fact, we checked it theoretically with the complex-scaling method, and
 found that the calculated resonant energy (width) is almost the same as the one
 with the bound state approximation (less than 100 keV, which is the numerical uncertainty 
 in the present calculation).
Thus, we use the $3^-$ wave function under the bound state approximation
 to study the characteristics of the state.
 
\begin{figure}
\begin{center}
\includegraphics*[scale=0.5,clip]{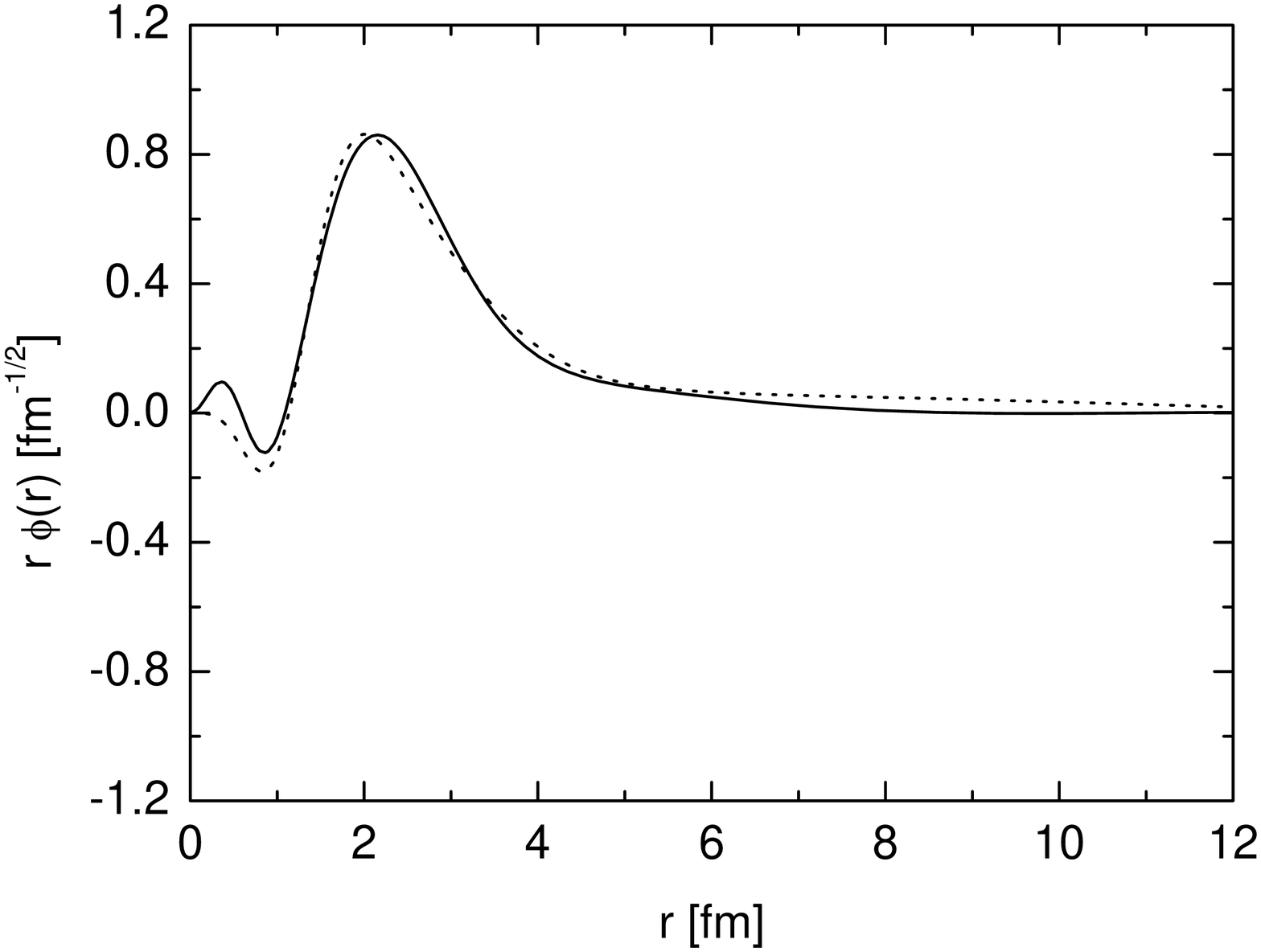}
\caption{
Radial parts of the single $\alpha$ orbits, $P_1$ (solid line)
 and $F_1$ (dotted), in the $3^-_1$ state.
Note that all of the radial parts in figures are multiplied by $r$.
}
\label{fig:9}
\end{center}
\end{figure}

The calculated nuclear radius for the $3^-$ state is 2.95 fm, the value of which
 is larger than that for the ground state ($0^+_1$), while it is smaller than 
 that for the $0^+_2$ state (see Table~\ref{tab:1}).
This suggests that the structure of the $3^-$ state is intermediate between the
 shell-model-like compact structure ($0^+_1$) and the dilute $3\alpha$ structure ($0^+_2$).
The occupation probability of the single-$\alpha$ orbits for the state are shown
 in Table~\ref{tab:3}:~$44.7 \%$ for $P_1$-orbit and $27.9 \%$ for $F_1$-orbit.
Although the concentration of the single orbit $P_1$ amounts to about $50 \%$,
 the radial behavior of the single-$\alpha$ orbit in Fig.~\ref{fig:9} has two nodes
 in the inner region.
However, the amplitude of the inner oscillations is significantly 
 smaller than that for the ground state in Fig.~\ref{fig:3}(a).
The small oscillations indicate the weak Pauli-blocking effect, and thus, we can
 see the precursor of the $3\alpha$ condensate state, although the $3^-$ state 
 is not an ideal rotating dilute $3\alpha$ condensate.

\subsection{$1^-_1$ state}

The experimental width of the $1^-_1$ state at $E_{3\alpha}^{exp}$=3.57 MeV
 is as small as $\Gamma$=315 keV.~\cite{Ajzenberg90}~
This means that the bound state approximation is good to describe the state.
In fact, the calculated energy of the $1^-_1$ state under the bound state approximation
 is $E_{3\alpha}$=3.11 MeV, which is quite similar to that with the complex scaling 
 method ($E_{3\alpha}$=3.1 MeV and $\Gamma$=0.1 MeV) and
 in good agreement with the experimental value.
 
\begin{figure}
\begin{center}
\includegraphics*[scale=0.5,clip]{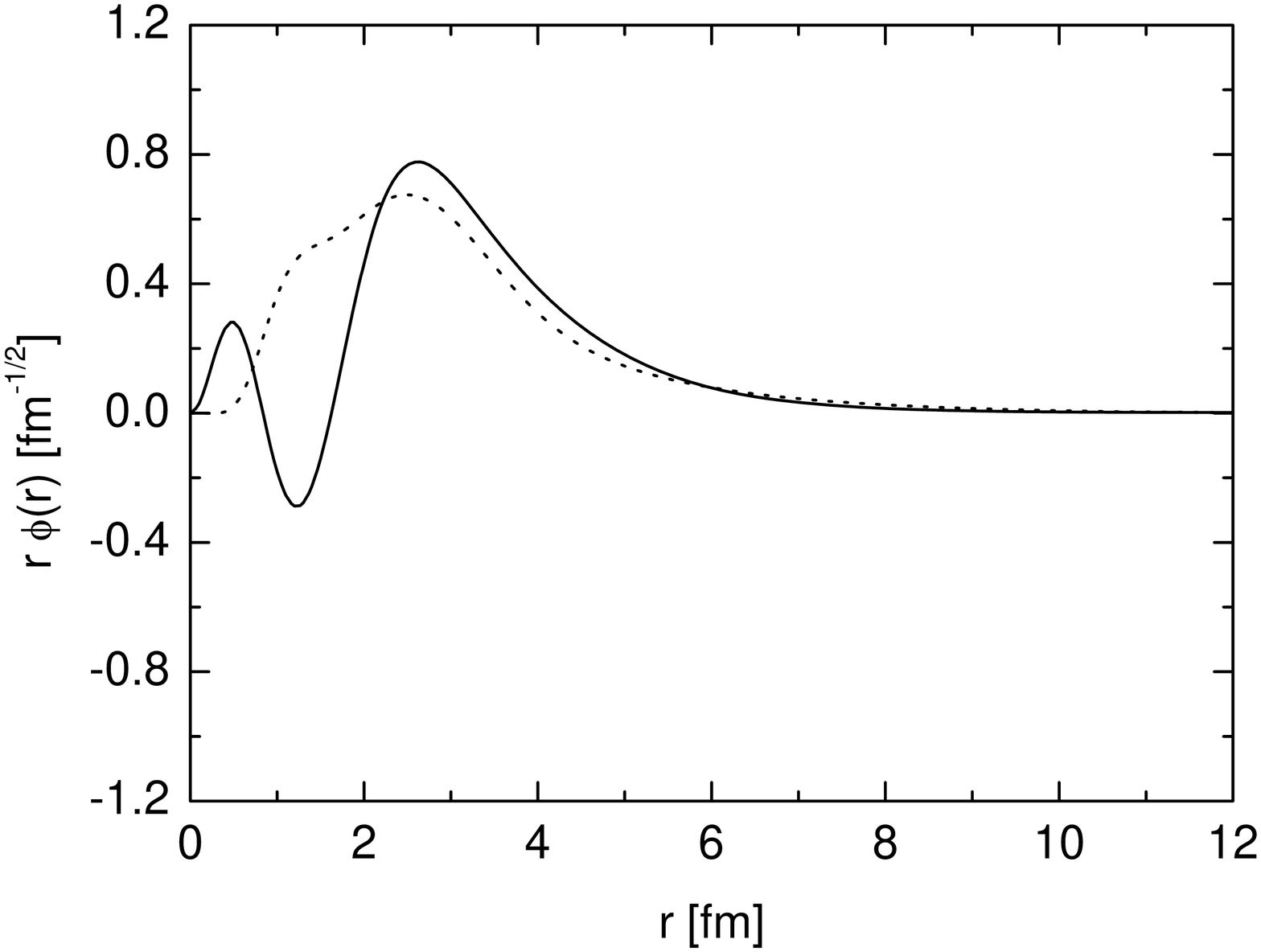}
\caption{
Radial parts of the single $\alpha$ orbits, $P_1$ (solid line)
 and $F_1$ (dotted), in the $1^-_1$ state.
Note that all of the radial parts in figures are multiplied by $r$.
}
\label{fig:10}
\end{center}
\end{figure}

The calculated nuclear radius, 3.32 fm, is larger than that of the ground state 
 (2.44 fm) and the $3^-_1$ state (2.95 fm) but is smaller than that of the $0^+_2$ one (4.3 fm).
The occupation probabilities of the $\alpha$ particles in the $1^-_1$ state are shown 
 in Table~\ref{tab:3}:~$35~\%$ for $P_1$ orbit and $16~\%$ for $F_1$ orbit.
Thus, there is no concentration of the occupation probability to a single orbit
 like the $0^+_2$ and $2^+_2$ states.
Since the $\alpha$ particles in the $1^-_1$ state are distributed over in several orbits, 
 the state is not of the dilute $\alpha$-condensate type.
Figure~\ref{fig:10} shows the radial behavior of the $P_1$ and $F_1$ orbits in the $1^-_1$ state.
The $P_1$ orbit has two nodes in the inner region, the behavior
 of which is rather similar to the $2P$ harmonic oscillator wave function.
However, the $F_1$ orbit has a $F$-wave Gaussian-type behavior.
(Exactly speaking, the orbit has one node at the vicinity of the origin, which can not
 be seen in Fig.~\ref{fig:10}.)~ 
Also we see the oscillatory behavior of the $F_1$ orbit for $ 0 < r < 2 $ fm,
 similar to the one of the $S_1$ orbit in the $0^+_2$ state in Fig.~\ref{fig:4}(b). 
These interesting behaviors of the $F_1$ orbit indicate some signal of 
 the dilute $\alpha$ condensation, reflecting the relatively large nuclear 
 radius (3.32 fm) for the $1^-_1$ state.

\subsection{Structure change of the $0^+$ state with nuclear radius}

In Sec.~IIIA, we found that the $0^+_2$ state has a dilute $3\alpha$ structure
 characterized by the nuclear radius as large as about 4.3 fm, in which
 the $\alpha$ particle occupies in the single orbit ($S_1$) with about $70~\%$ probability, 
 and the radial behavior of the $S_1$ orbit is similar to the $S$-wave Gaussian wave 
 function with a very long tail.
On the other hand, the $0^+_1$ state has a compact structure with 
 a nuclear radius of 2.44 fm, where the occupation probabilities of the $\alpha$ 
 particles spread out over the $S$, $D$ and $G$ orbits, amounting to about $30~\%$, each.
The feature is much in contrast with that of the $0^+_2$ state.   
The nuclear radius or density of $^{12}$C seems to have a close relation with making the
 compact structure and the dilute $3\alpha$ structure in the $^{12}$C $0^+$ state.  
Thus, it is very interesting to see the structure change of the $0^+$ state of $^{12}$C
 by taking the nuclear radius (or density) as parameter.  
The dependence of the occupation probabilities and radial behaviors 
 of the single $\alpha$-particle orbits in the $0^+$ state on its nuclear radius
 is investigated with the use of the simple framework given in the latter part 
 of Sec.~IIC (see Eqs.~(\ref{eigen_eq_Pauli_operator})$\sim$(\ref{eigen_eq_density_rms})).

\begin{figure}
\begin{center}
\includegraphics*[scale=0.5,clip]{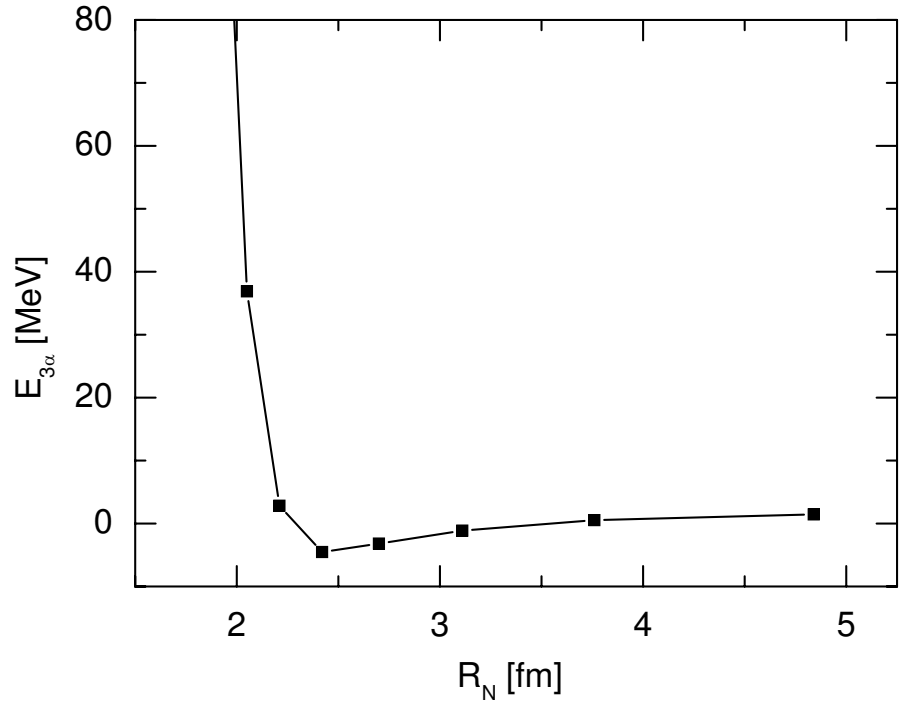}
\caption{
Dependence of the energy of the $^{12}$C($0^+$) state measured 
from the $3\alpha$ threshold on its nuclear radius.
}
\label{fig:11}
\end{center}
\end{figure}

Figure~\ref{fig:11} shows the dependence of the energy of $^{12}$C measured from 
 the $3\alpha$ threshold on the nuclear radius $R_N$, 2.20 fm $ \leq R_N \leq 4.86$ fm,
 corresponding to a nuclear density 0.15 $\leq \rho/\rho_0 \leq$ 1.6 ($\rho_0$ denotes
 the normal density).
The energy minimum point appears around $R_N\sim$2.4 fm, corresponding to 
 the normal density region.
We see the strong repulsion in the region of $R_N < 2.2$~fm, due to the
 kinetic-energy effect and Pauli-blocking effect, while the almost flat
 region appears at $R_N > 4$~fm and the energy is positive and small, less than 1 MeV
 with respect to the $3\alpha$ threshold.

\begin{figure}
\begin{center}
\includegraphics*[scale=0.5,clip]{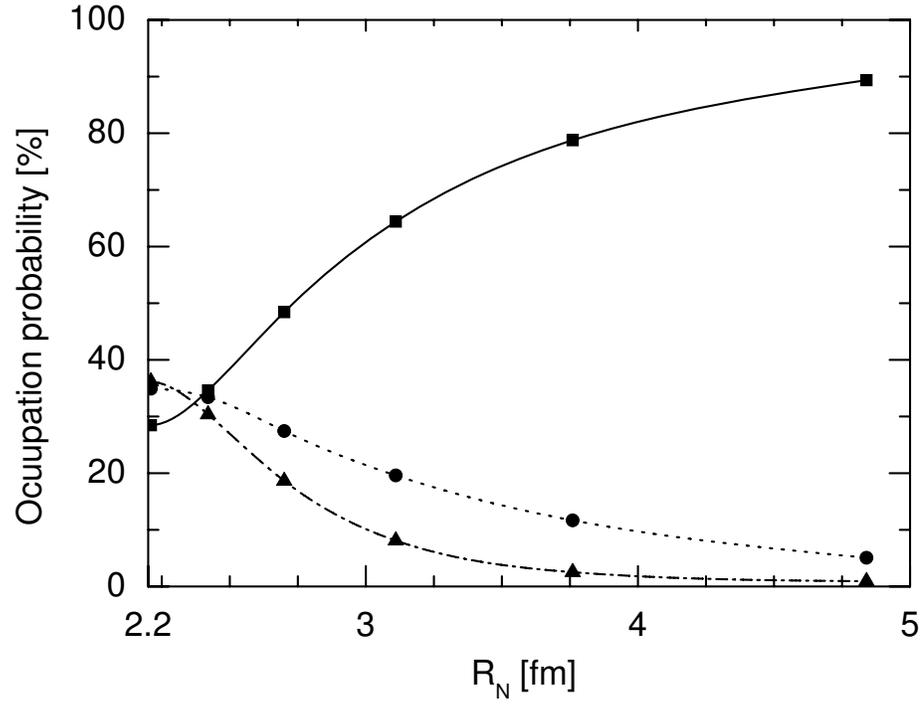}
\caption{
Dependence of the occupation probabilities of the single-$\alpha$ orbits
($S_1$-, $D_1$- and $D_1$-orbits) in the $^{12}$C($0^+$) state
on its nuclear radius.
The solid (dotted and dot-dashed) line corresponds
to the $S_1$-orbit ($D_1$- and $G_1$ orbits, respectively).
}
\label{fig:12}
\end{center}
\end{figure}

The occupation probabilities of the single-$\alpha$ orbits ($S_1$-, $D_1$-, and 
 $G_1$-orbits) are shown in Fig.~\ref{fig:12} with respect to the nuclear radius, 
 where $L_k$ denotes the $k$-th orbit for the $L$ wave.
In the region of $R_N=2.2\sim2.4$~fm (normal density region), 
 the occupation probabilities of the $\alpha$ particles spread out 
 over the $S$, $D$ and $G$ orbits, amounting to about $30~\%$, each.
This feature is almost the same as that of the $0^+_1$ state obtained by the $3\alpha$
 OCM calculation, the nuclear radius of which is 2.43 fm (see Sec.~IIIA).
Figure~\ref{fig:13} shows the radial behavior of the single-$\alpha$ orbit, $S_1$,  
  with respect to the nuclear radius.
The $S_1$-orbit at $R_N\sim2.42$~fm (Fig.~\ref{fig:13}(a)) has two nodes and 
 the radial behavior is of the $2S$ harmonic oscillator wave function (howf) type, 
 the result of which is almost the same as that of the $0^+_1$ state obtained 
 by the $3\alpha$ OCM calculation (see Fig.~\ref{fig:4}(a)).   
Thus, the wave function with $R_N\sim2.4$~fm has the SU(3)$[f](\lambda\nu)=[444](04)$
 character (see Eq.~(\ref{SU3_0+})).

Increasing the nuclear radius from $R_N=2.42$~fm, the occupation probability 
 of the single-$\alpha$ orbits concentrates gradually on a single orbit ($S_1$), 
 and it amounts to be about 90~\% at $R_N$=4.84 fm ($\rho/\rho_0$=0.14)
 in the present calculation (see Fig.~\ref{fig:12}).
The radial behaviors of the $S_1$ orbit with $R_N$=2.42, 2.70, 3.11 and 4.84 fm 
 are demonstrated in Figs.~\ref{fig:13}(a), (b), (c) and (d), respectively.
We can see that increasing the nuclear radius, the internal oscillation observed 
 in the $S_1$ orbit with $R_N$=2.42 fm is gradually disappearing and, finally, 
 the $2S$-type radial behavior transits to the zero-node long-ranged 
 $S$-wave type (Gaussian) with the occupation probability of about 90~\%, 
 approaching an ideal dilute $\alpha$ condensate.
The reason of why only the $S$ wave survives in the case of increasing 
 the nuclear radius is due to the fact that the centrifugal barrier 
 is not at work for the $S$-wave $\alpha$ orbit.
The $S$-wave $\alpha$ particles, thus, 
 can move in a nucleus with a given large nuclear radius, although they
 are confined by the Coulomb potential barrier produced self-consistently~\cite{Yamada04}.    
According to the results of the $3\alpha$ OCM calculation (see Sec.~IIIA),
 the $\alpha$ particle in the $0^+_2$ state ($R_N$=4.3 fm) is occupied 
 in the single orbit ($S_1$) with about 70~\% probability, the radial behavior of which 
 is similar to the $S$-wave Gaussian wave function with a very long tail.
Their results are consistent with those in Figs.~\ref{fig:12} and \ref{fig:13}.

\newpage
\section{Summary}

In this work we have investigated the bosonic properties such as single-$\alpha$ particle
 orbits and occupation numbers in the $J^\pi$=$0^+$, $2^+$, $1^-$, and $3^-$ states 
 of $^{12}$C around the $3\alpha$ threshold within the framework of the $3\alpha$ OCM 
 (orthogonality condition model).
The $3\alpha$ OCM equation is based on the equation of motion for 
 the $N\alpha$ bosons derived from the microscopic $N\alpha$ cluster model theory.
The experimental energy spectra for $0^+_1$, $0^+_2$, $2^+_2$, $1^-_1$,
 and $3^-_1$ are reproduced well with the $3\alpha$ OCM.

The main results to be emphasized here are as follows.

(1)~The $0^+_2$ state at $E^{exp}_{3\alpha}$=0.38 MeV has a dilute $3\alpha$ structure
 characterized by the nuclear radius as large as about 4.3 fm.
The analysis of the single-$\alpha$ orbits and occupation probabilities
 for the dilute state shows that the $\alpha$ particle is occupied in a single
 orbit ($S_1$) with about $70~\%$ probability, and the radial behavior 
 of the $S_1$ orbit is similar to the $S$-wave Gaussian wave function with a very long tail.
The momentum distribution of the $\alpha$ particle illustrates the $\delta$-function like 
 behavior, similar to the momentum distribution of a dilute neutral atomic
 condensate states at very low temperature, a feature which eventually can be measured
 experimentally.
These results give more theoretical evidence that the $0^+_2$ state is 
 a dilute $3\alpha$ condensate.
On the other hand, the $0^+_1$ state has a compact structure with 
 a nuclear radius of 2.44 fm. 
The occupation probabilities of the $\alpha$ particles spread out over 
 the $S$, $D$ and $G$ orbits, amounting to about $30~\%$, each,
 the results of which comes from the fact that the $0^+_1$ state is characterized 
 by the nuclear SU(3) wave function, $[f](\lambda\mu)=[444](04)$.
The feature is much in contrast with that of the $0^+_2$ state.   

(2)~In order to understand further the characteristic structure of the two $0^+$ states,
 we have studied the single-$\alpha$ orbital behavior
 in the $^{12}$C($0^+$) state by taking the nuclear radius $R_N$ (or density $\rho/\rho_0$) 
 as parameter, 2.42 $\leq R_N \leq$ 4.84 fm, (0.15 $\leq \rho/\rho_0 \leq$ 1.2),
 where $\rho_0$ denotes the normal density).
We found that the single-$\alpha$ orbits in the $^{12}$C($0^+$) state
 are smoothly changed with the nuclear radius $R_N$, and their behavior is classified 
 into the following three types, depending on the value of $R_N$:~
i)~at $R_N\sim2.4$~fm ($\rho/\rho_0\sim$ 1), 
 we have two-nodal $S$-orbit ($2S$), one-nodal $D$-orbit ($1D$) and zero-nodal
 $G$-orbit ($0G$) with about $30 \%$ occupation probability, each, 
 characterized by a nuclear SU(3) wave function, 
ii)~increasing the nuclear radius from $R_N\sim2.4$ fm, the occupation
 probability of the single-$\alpha$ orbits concentrates gradually on a single $S$-orbit, 
 in which the two-nodal behavior is disapperaing, and then,  
iii)~at $R_N\sim 4$ fm ($\rho/\rho_0\sim$ 0.2), there appears a dominant zero-nodal 
 Gaussian ($0S$-type) orbit with a very long tail, the radial behavior of 
 which is similar to that of the $0^+_2$ state in $^{12}$C as mentioned above.
The structure change is caused mainly by the Pauli-blocking effect, the strength
 of which depends dominantly on the nuclear radius $R_N$ in the present framework.

(3)~The structure of the $2^+_2$ state at $E^{exp}_{3\alpha}$=$2.6\pm0.3$ MeV 
 with $\Gamma=1.0\pm0.3$ MeV was studied with the present $3\alpha$ OCM
 and the complex-scaling method.
We found that the $2^+_2$ resonant state appears at $E_{3\alpha}$=2.3 MeV with 
 $\Gamma=$1.0 MeV, in agreement with the experimental data, and the
 calculated nuclear radius is 4.3 fm, similar to that of the $0^+_2$ state.
The $2^+_2$ wave function obtained with the $3\alpha$ OCM was used to
 study the bosonic properties of the state.
It was found that the occupation probability of the $\alpha$ particle concentrates
 only on the $D_1$ orbit, amounting to be as large as about $80~\%$, and
 the radial behavior is of the $D$-wave Gaussian type with long tail.
The characteristics of the boson properties in $2^+_2$ is quite similar
 to those in $0^+_2$ at $E^{exp}_{3\alpha}$=0.38 MeV.
Thus, the $2^+_2$ state has the dilute $3\alpha$-condensate-like structure.  
On the other hand, the $2^+_1$ state has a compact structure with 
 the nuclear radius, 2.44 fm, like the ground state. 
The occupation probabilities of the $\alpha$ particles spread out over 
 the $D$ and $G$ orbits, amounting to about $56~\%$ and $33~\%$, 
 respectively, reflecting the SU(3) character of the $2^+_1$ state.   

(4)~We investigated the $\alpha$ bosonic properties of the negative parity states,
  $1^-_1$ at $E^{exp}_{3\alpha}$=3.57 MeV and $3^-_1$ at $E^{exp}_{3\alpha}$=2.37 MeV.
Their nuclear radii are 3.32 and 2.95 fm, respectively, which are larger than
 that of the ground state ($0^+_1$) but smaller than that of $0^+_2$.
The calculated occupation probabilities of the $\alpha$ particles in those states 
 show that there is no concentration on a single $\alpha$ orbit like in the $0^+_2$
 and $2^+_2$ states.  
The results indicates that the $1^-$ and $3^-_1$ states are not 
 of the dilute $3\alpha$ condensate.
The radial behavior of the $P$- and $F$-wave single $\alpha$ orbits, however,
 suggests that small components of the $3\alpha$ condensation
 exist even in the negative parity states, which is reflected 
 by their relatively large nuclear radii.

\section*{Acknowledgments}
We acknowledge helpful discussions with H.~Horiuchi, K.~Ikeda,
 G.~R\"opke, Y.~Suzuki, and A.~Tohsaki.

\clearpage

\clearpage
\begin{table}
\caption{
Calculated energies ($E_{3\alpha}$) and nuclear radii ($\sqrt{{\langle r^2_N \rangle}}$)
 for the $0^+$, $2^+$, $3^-$ and $1^-$
 states of $^{12}$C together with the $\alpha$-$\alpha$ and $\alpha$-$2\alpha$
 rms distances ($\sqrt{{\langle r^2_{\alpha\alpha}\rangle}}$ and
 $\sqrt{{\langle r^2_{\alpha-2\alpha}\rangle}}$).
The energy $E_{3\alpha}$ is one measured from the $3\alpha$ threshold.
The values in parentheses denote the experimental ones.
All energies and nuclear radii (rms distances) are given in units of MeV and fm, respectively.
}
\label{tab:1}
\begin{center}
\begin{tabular}{cccccc}
\hline
\hline
\hspace{10mm}$J^\pi$\hspace{8mm} 
     & \hspace{5mm}{$E_{3\alpha}^{cal}$}\hspace{5mm} 
     & \hspace{5mm}{$(E_{3\alpha}^{exp})$}\hspace{5mm}
     & \hspace{5mm}{$\sqrt{{\langle r^2_N \rangle}}$}\hspace{5mm} 
     & \hspace{5mm}{$\sqrt{{\langle r^2_{\alpha\alpha}\rangle}}$}\hspace{5mm}
     & \hspace{5mm}{$\sqrt{{\langle r^2_{\alpha-2\alpha}\rangle}}$}\hspace{5mm}\\
\hline
 $0^+_1$ & $-7.27~$ & $(-7.27)$ & 2.44~ & 3.02 & 2.61 \\ 
 $0^+_2$ & $~~0.85~$  & $(~~0.38)$   & 4.31~ & 6.84 & 5.93 \\[1mm]
 $2^+_1$ & $-5.28~$ & $(-2.83)$ & 2.45~ & 2.94 & 2.55 \\ 
 $2^+_2$ & $~~2.43^*$  & $(~~2.6~)$   & 6.12$^*$ & 10.2 & 8.80 \\[1mm] 
 $3^-_1$ & $~~1.52~$ & $(~~2.37)$ & 2.96~ & 4.10 & 3.56 \\[1mm] 
 $1^-_1$ & $~~3.11~$& $(~~3.57)$ & 3.32~ & 4.87 & 4.23 \\[1mm] 
\hline
\hline
\end{tabular}
\end{center}
\footnotetext{$^*$According to the complex-scaling method, the resonant energy and width
of the $2^+_2$ state are $E_{3\alpha}=2.3$ MeV and $\Gamma=1.0$ MeV, respectively, 
with $\sqrt{{\langle r^2_N \rangle}}$=4.3 fm. See text.}  
\end{table}

\clearpage
\begin{table}
\caption{
Occupation numbers of the $k$-th $\alpha$-orbits
 with $S$-, $D$- and $G$-waves for the $0^+$ and $2^+$ states of $^{12}$C obtained 
 by diagonalizing the one-body density matrix in Eq.~(\ref{one_body_density}).
The values in parentheses denote the occupation probabilities.
}
\label{tab:2}
\begin{center}
\begin{tabular}{cccccc}
\hline
\hline
\hspace{10mm}$J^\pi$\hspace{10mm}
     & $k$
     & \hspace{5mm}{$S$}\hspace{5mm} 
     & \hspace{5mm}{$D$}\hspace{5mm}
     & \hspace{5mm}{$G$}\hspace{5mm} \\
\hline
 $0^+_1$ & $1$ & \hspace{5mm}{1.05 ($35.0~\%$)}\hspace{5mm} & \hspace{5mm}{1.06 ($35.3~\%$)}\hspace{5mm}  & \hspace{5mm}{0.82 ($27.3~\%$)}\hspace{5mm} \\  
         & $2$ & 0.01 ($~0.3~\%$) & 0.01 ($~0.0~\%$) & 0.00 ($~0.0~\%$) \\  
         & $3$ & 0.00 ($~0.0~\%$) & 0.01 ($~0.0~\%$) & 0.00 ($~0.0~\%$) \\  
         & $\cdots$ & $\cdots$ & $\cdots$ & $\cdots$ \\  
         & total & 1.07 ($35.6~\%$) & 1.07 ($35.6~\%$) & 0.82 ($27.3~\%$) \\
\hline
 $0^+_2$ & $1$ & \hspace{5mm}{2.16 ($72.0~\%$)}\hspace{5mm} & \hspace{5mm}{0.19 ($~6.3~\%$)}\hspace{5mm}  & \hspace{5mm}{0.08 ($~2.7~\%$)}\hspace{5mm} \\  
         & $2$ & 0.20 ($~6.7~\%$) & 0.06 ($~2.0~\%$) & 0.06 ($~2.0~\%$) \\  
         & $3$ & 0.02 ($~0.7~\%$) & 0.02 ($~0.7~\%$) & 0.01 ($~0.3~\%$) \\  
         & $\cdots$ & $\cdots$ & $\cdots$ & $\cdots$ \\  
         & total & 2.38 ($79.3~\%$) & 0.29 ($~1.0~\%$) & 0.16 ($~5.3~\%$) \\
\hline
 $2^+_1$ & $1$ & \hspace{5mm}{0.25 ($~8.5~\%$)}\hspace{5mm} & \hspace{5mm}{1.69 ($56.2~\%$)}\hspace{5mm}  & \hspace{5mm}{1.00 ($33.3~\%$)}\hspace{5mm} \\  
         & $2$ & 0.00 ($~0.0~\%$) & 0.02 ($~0.7~\%$) & 0.00 ($~0.0~\%$) \\  
         & $3$ & 0.00 ($~0.0~\%$) & 0.00 ($~0.0~\%$) & 0.00 ($~0.0~\%$) \\  
         & $\cdots$ & $\cdots$ & $\cdots$ & $\cdots$ \\  
         & total & 0.26 ($~8.7~\%$) & 1.71 ($57.0~\%$) & 1.00 ($33.3~\%$) \\
\hline
 $2^+_2$ & $1$ & \hspace{5mm}{0.31 ($10.3~\%$)}\hspace{5mm} & \hspace{5mm}{2.50 ($83.3~\%$)}\hspace{5mm}  & \hspace{5mm}{0.05 ($~1.7~\%$)}\hspace{5mm} \\  
         & $2$ & 0.02 ($~0.7~\%$) & 0.03 ($~1.0~\%$) & 0.00 ($~0.0~\%$) \\  
         & $3$ & 0.00 ($~0.0~\%$) & 0.01 ($~0.3~\%$) & 0.00 ($~0.0~\%$) \\  
         & $\cdots$ & $\cdots$ & $\cdots$ & $\cdots$ \\  
         & total & 0.33 ($11.0~\%$) & 2.56 ($85.3~\%$) & 0.06 ($~2.0~\%$) \\
\hline
\hline
\end{tabular}
\end{center}
\end{table}

\clearpage
\begin{table}
\caption{
Occupation numbers of the $k$-th $\alpha$-orbits
 with $P$- and $F$-waves for the $3^-$ and $1^-$ states of $^{12}$C obtained 
 by diagonalizing the one-body density matrix in Eq.~(\ref{one_body_density}).
The values in parentheses denote the occupation probabilities.
}
\label{tab:3}
\begin{center}
\begin{tabular}{ccccc}
\hline
\hline
\hspace{10mm}$J^\pi$\hspace{10mm}
     & $k$
     & \hspace{5mm}{$P$}\hspace{5mm} 
     & \hspace{5mm}{$F$}\hspace{5mm} \\
\hline
 $3^-_1$ & $1$ & \hspace{5mm}{1.34 ($44.7~\%$)}\hspace{5mm} & \hspace{5mm}{0.84 ($27.9~\%$)}\hspace{5mm} \\  
         & $2$ & 0.12 ($~4.0~\%$) & 0.23 ($~7.5~\%$) \\  
         & $3$ & 0.06 ($~1.9~\%$) & 0.02 ($~0.8~\%$) \\  
         & $\cdots$ & $\cdots$ & $\cdots$ \\  
         & total & 1.54 ($51.4~\%$) & 1.09 ($36.4~\%$) \\
\hline
 $1^-_1$ & $1$ & \hspace{5mm}{1.06 ($35.3~\%$)}\hspace{5mm} & \hspace{5mm}{0.47 ($15.8~\%$)}\hspace{5mm} \\  
         & $2$ & 0.53 ($17.8~\%$) & 0.26 ($~8.6~\%$) \\  
         & $3$ & 0.08 ($~2.6~\%$)  & 0.08 ($~2.6~\%$) \\  
         & $\cdots$ & $\cdots$ & $\cdots$ \\  
         & total & 1.75 ($58.5~\%$) & 0.84 ($28.1~\%$) \\
\hline
\hline
\end{tabular}
\end{center}
\end{table}

\clearpage
 
\clearpage

\clearpage

\clearpage

\clearpage

\clearpage

\clearpage

\clearpage

\clearpage

\clearpage

\clearpage

\clearpage

\clearpage
\begin{figure}
\begin{center}
\includegraphics*[scale=0.5,clip]{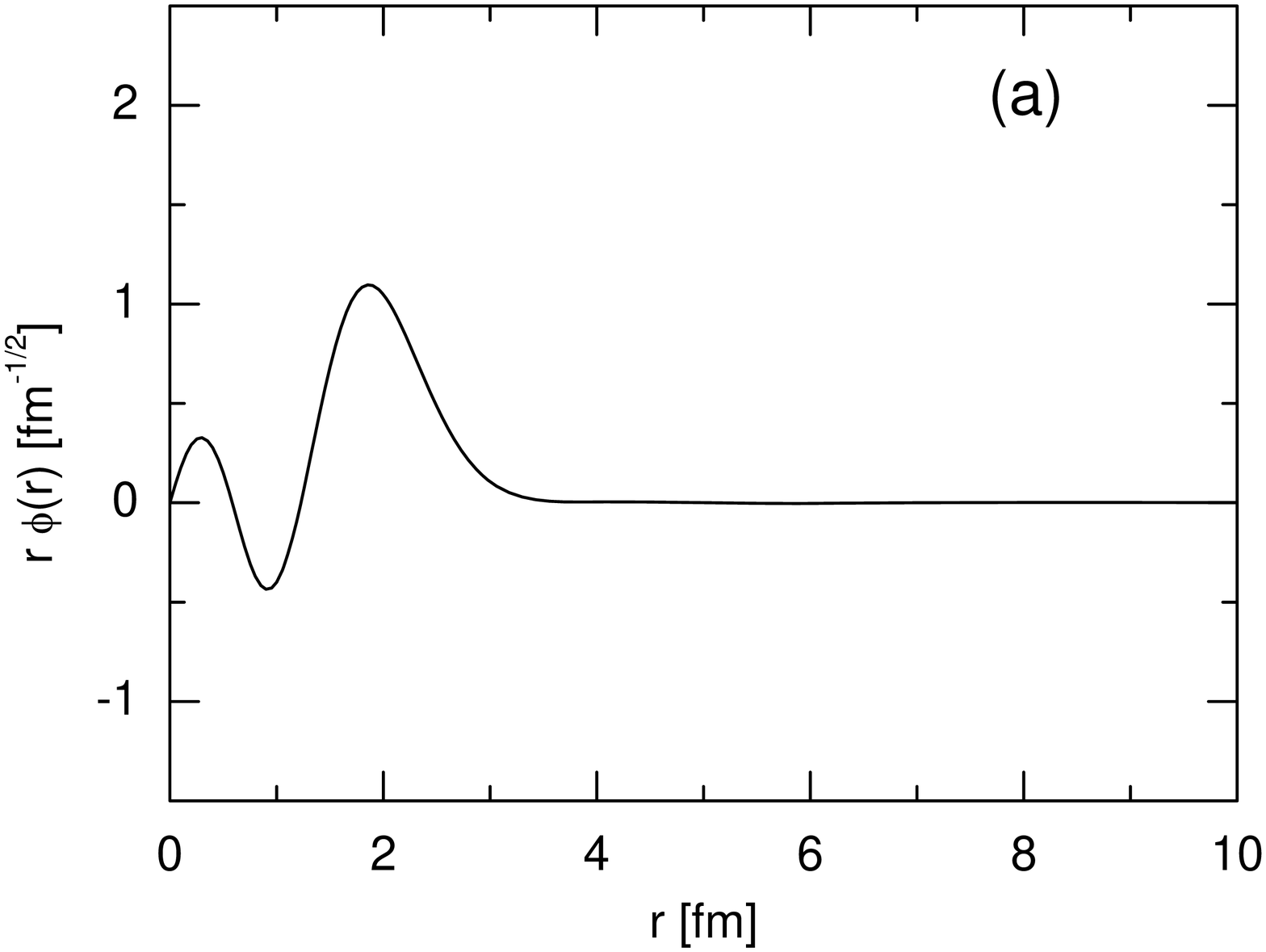}
\includegraphics*[scale=0.5,clip]{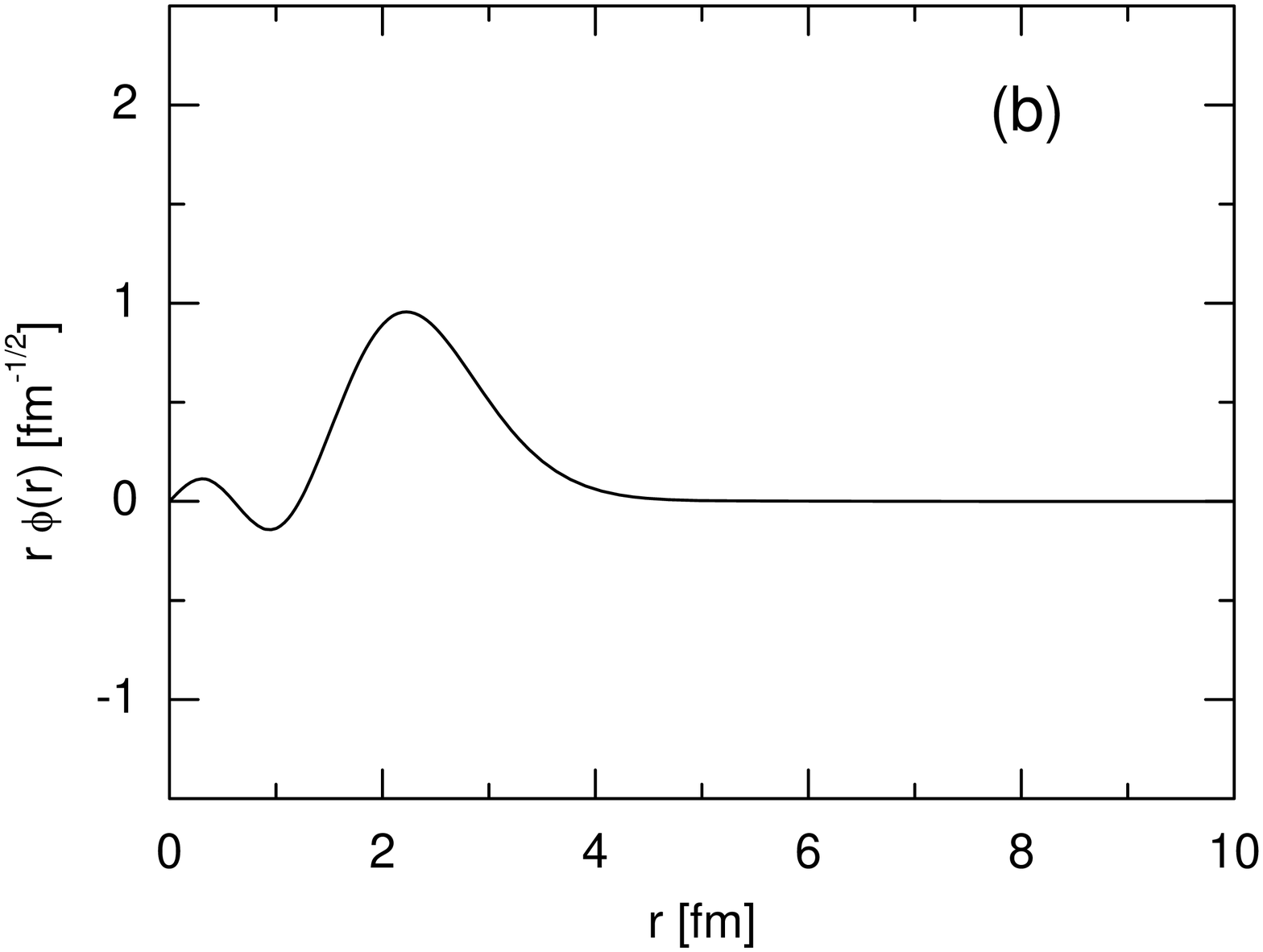}
\end{center}
\end{figure}
\begin{figure}
\begin{center}
\includegraphics*[scale=0.5,clip]{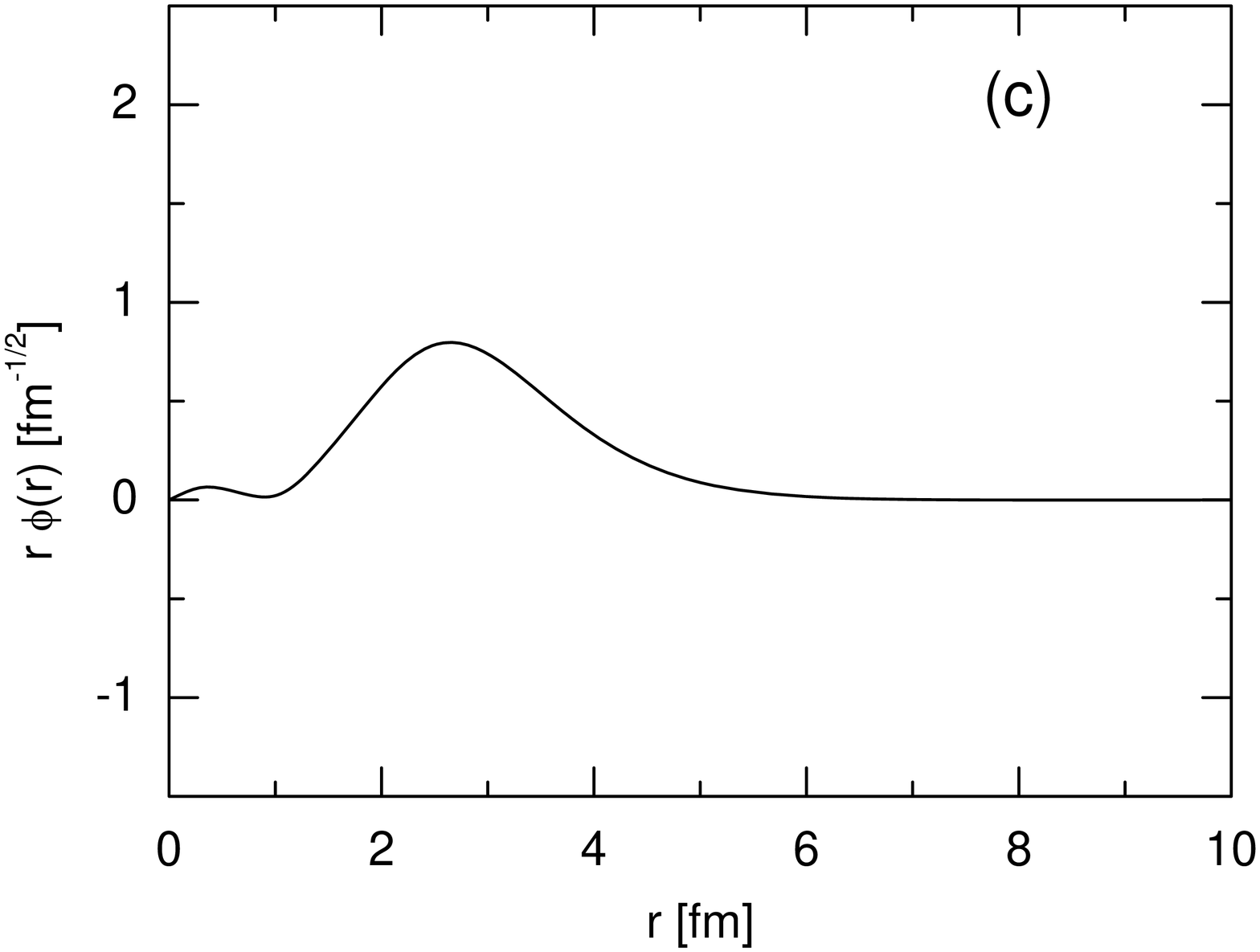}
\includegraphics*[scale=0.5,clip]{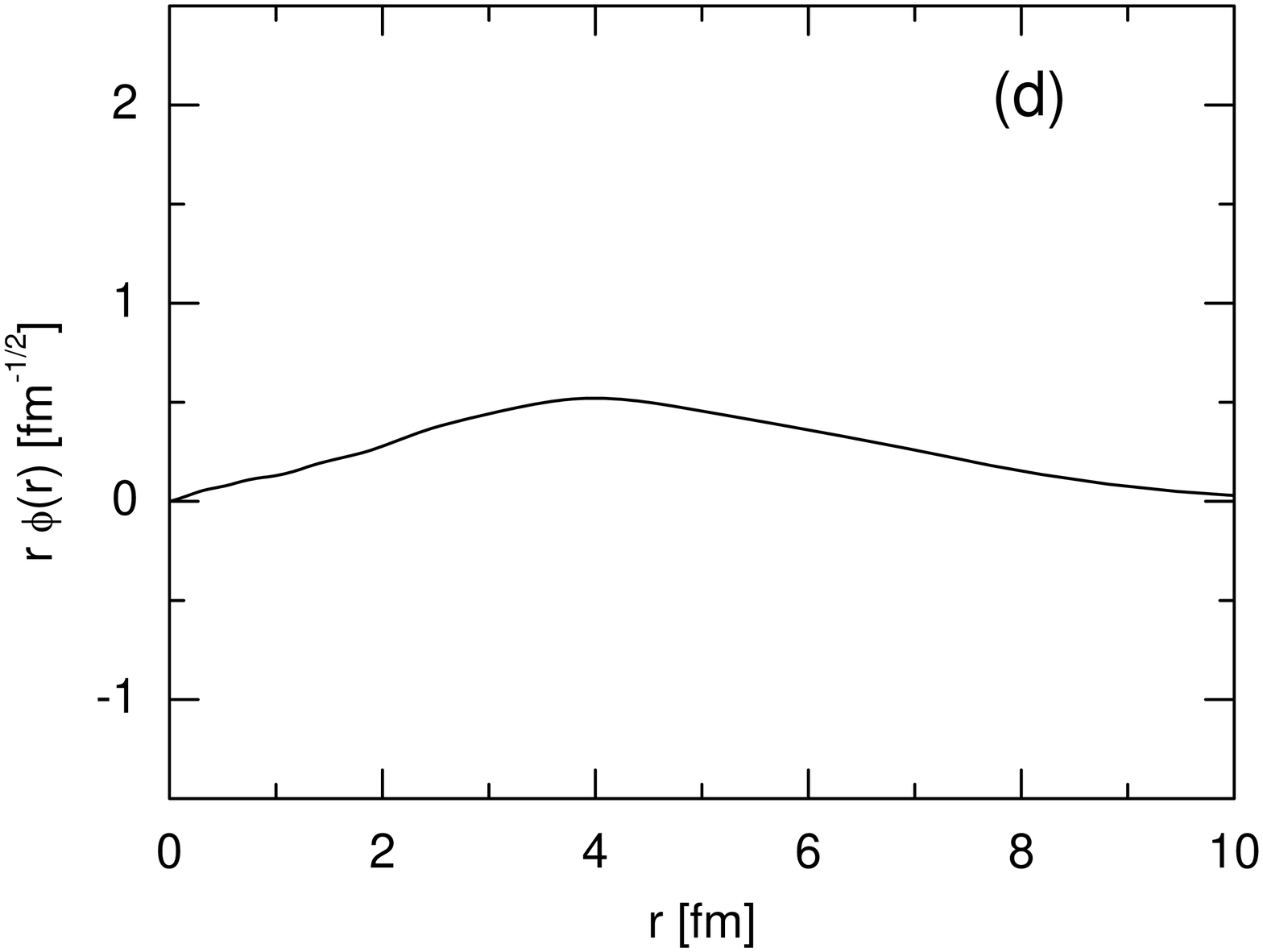}
\caption{
Radial behaviors of the $S_1$-orbit in the $^{12}$C($0^+$) state
with (a)~$R_N$=2.42 fm, (b)~$R_N$=2.70 fm, (c)~$R_N$=3.11 fm, 
and (d)~$R_N$=4.84 fm,
where $R_N$ denotes the nuclear radius of the $^{12}$C($0^+$) state.
}
\label{fig:13}
\end{center}
\end{figure}
 

\begin{thebibliography}{99}

\bibitem{Wildermuth77}
   K.~Wildermuth and Y.C.~Tang, {\it A Unified Theory of the Nucleus} 
   (Vieweg, Braunschweig, Germany, 1977).

\bibitem{Brink66}
   D.M.~Brink, in {\it Many-Body Description of Nuclear Structure and Reactions}, 
   Proceedings of the International School of Physics "Enrico Fermi", Course 36,
   edited by C.~Bloch (Academic Press, New York, 1966).

\bibitem{Bertsch71}
   G.F. Bertsch and W. Bertozzi, Nucl.\ Phys.\ {\bf A165}, 199 (1971).

\bibitem{Fujiwara80}
   Y.~Fujiwara, H.~Horiuchi, K.~Ikeda, M.~Kamimura, K.~Kato, Y.~Suzuki, and E.~Uegaki,
   Prog.\ Theor.\ Phys.\ Suppl.\ No.\ 68, 29 (1980). 

\bibitem{Horiuchi86}
   H.~Horiuchi and K.~Ikeda, {\it Cluster Model of the Nucleus},
   International Review of Nuclear Physics, World Scientific Publishing Co., {\bf 4}, 1 (1986).
   
\bibitem{Uegaki77}
   E.~Uegaki, S.~Okabe, Y.~Abe, and H.~Tanaka, Prog.\ Theor.\ Phys.\ {\bf 57}, 1262 (1977).\\
   E.~Uegaki, Y.~Abe, S.~Okabe, and H.~Tanaka, Prog.\ Theor.\ Phys.\ {\bf 59}, 1031 (1978); 
   {\bf 62}, 1621 (1979).   

\bibitem{Fukushima77}
   Y. Fukushima and M. Kamimura, Proc.~Int.~Conf.~on Nuclear Structure, Tokyo, 1977, ed.~ T.~Marumori 	 
   (Suppl.~of J.~Phys.~Soc.~Japan, Vol.44, 1978), p.~225.\\
  M.~Kamimura, Nucl. Phys. {\bf A 351}, 456 (1981). 	 

\bibitem{Dalfovo99}
   F.~Dalfovo, S.~Giorgini, L.P.~Pitaevskii, and S.~Stringari, Rev.\ Mod.\ Phys.\ 
   {\bf 71}, 463 (1999).

\bibitem{Ropke98}
   G.~R\"opke, A.~Schnell, P.~Schuck, and P.~Nozieres, 
   Phys.\ Rev.\ Lett.\ {\bf 80}, 3177 (1998).     

\bibitem{Beyer00}
   M.~Beyer, S.A.~Sofianos, C.~Kuhrts,  G.~R\"opke, and P.~Schuck, 
   Phys.\ Lett.\ B {\bf 80}, 247 (2000).     

\bibitem{Tohsaki01} 
   A.~Tohsaki, H.~Horiuchi, P.~Schuck and G.~R\"opke, 
   Phys.\ Rev.\ Lett.\ {\bf 87}, 192501 (2001). 

\bibitem{Funaki02}
   Y.~Funaki, H.~Horiuchi, A.~Tohsaki, P.~Schuck and G.~R\"opke, 
   Prog.\ Theor.\ Phys.\ {\bf 108}, 297 (2002).

\bibitem{Yamada04}
  T.~Yamada and P.~Schuck, Phys.\ Rev.\ C\ {\bf 69}, 024309 (2004).   

\bibitem{Pitaevskii61}
   L.P.~Pitaevskii, Zh.\ Eksp.\ Theor.\ Fiz.\ {\bf 40}, 646 (1961) [Sov.\ Phys.\ JETP {\bf 13}, 451 (1961)].\\
   E.P.~Gross, Nuovo\ Cimento {\bf 20}, 454 (1961); J.\ Math.\ Phys.\ {\bf 4}, 195 (1963).

\bibitem{Itoh04}
   M.~Itoh et al., Nucl.~Phys.~{\bf A 738}, 268 (2004).

\bibitem{Funaki04}
   Y.~Funaki, H.~Horiuchi, A.~Tohsaki, P.~Schuck and G.~R\"opke, 
   arXiv:~nucl-th/0410097, 2004.
   
\bibitem{Kukulin77}
  V.I.~Kukulin and V.~M. Krasnopol'sky, J.~Phys. {\bf A 10}, 33 (1977).\\
  V.I.~Kukulin, V.M.~Krasnopol'sky, and M.~Miselkhi, Sov.~J.~Nucl.~Phys. {\bf 29}, 421 (1979).
    
\bibitem{Matsuura04}
   H.~Matsumura and Y.~Suzuki, Nucl.~Phys.~{\bf A 739}, 238 (2004).

\bibitem{Saito68}
   S.~Saito, Prog.\ Theor.\ Phys.\ {\bf 40}, 893 (1968); {\bf 41}, 705 (1969). 
   
\bibitem{Horiuchi74}
   H.~Horiuchi, Prog.\ Theor.\ Phys.\ {\bf 51}, 1266 (1974); {\bf 53}, 447 (1975).

\bibitem{Kato89}
   K.~Fukatsu, K.~Kato and H.~Tanaka, Prog.\ Theor.\ Phys.\ {\bf 81} (1989) 738.\\ 
   K.~Fukatsu and K.~Kato, Prog.\ Theor.\ Phys.\ {\bf 87}, 151 (1992).

\bibitem{Kurokawa04}
   C.~Kurokawa and K.~Kato, Nucl.~Phys.~{\bf A 738}, 455 (2004).

\bibitem{Kamimura88}
   M.~Kamimura, Phys.~Rev.~A~{\bf 38}, 621 (1988).\\
   H.~Kameyama, M.~Kamimura, and Y.~Fukushima, Phys.~Rev.~C~{\bf 40}, 974 (1989).\\
   E.~Hiyama, Y.~Kino, and M.~Kamimura, Prog.~Part.~Nucl.~Phys. {\bf 51}, 1 (2003).

\bibitem{Kukulin84}
   V.I.~Kukulin, V.M.~Krasnopol'sky, V.T.~Voronchev, and P.B.~Sazanov,
   Nucl.~Phys.~{\bf A417}, 128 (1984).

\bibitem{Ajzenberg90}
   F.~Ajzenberg-Selove, Nucl.\ Phys.\ {\bf A 506}, 1 (1990).

\bibitem{Kuruppa88}
   A.T.~Kruppa, R.G.~Lovas, and B.~Gyarmati, Phys.~Rev.~C {\bf37}, 383 (1988).\\
   M.~Homma, T.~Myo, and K.~Kato, Prog.~Theor.~Phys.~{\bf 97}, 561 (1996).

\end{thebibliography}
\end{document}